\documentclass{jfm}
\usepackage{graphicx}
\usepackage{epstopdf, epsfig}
\usepackage{amsmath}
\usepackage{color}

\usepackage{sistyle}
\shorttitle{Oscillatory liquid metal rotating convection}
\shortauthor{T. Vogt, S. Horn, and J.M. Aurnou}

\usepackage{overpic}
\usepackage{rotating}
\renewcommand{\vec}{\boldsymbol}

\usepackage[dvipsnames]{xcolor}
\newcommand{\SH}[1]{{\color{black}{#1}}}
\newcommand{\sus}[1]{{\color{black}{#1}}}
\newcommand{\JA}[1]{{\color{black}{#1}}}
\newcommand{\JMA}[1]{{\color{black}{#1}}}
\newcommand{\TV}[1]{{\color{black}{#1}}}

\title{\TV{Oscillatory thermal-inertial flows in liquid metal rotating convection}}
\author{Tobias Vogt\aff{1,2}
  \corresp{\email{t.vogt@hzdr.de}},
  Susanne Horn\aff{2,3},
 \and Jonathan M. Aurnou\aff{2}}

\affiliation{\aff{1}Institute of Fluid Dynamics, Helmholtz-Zentrum Dresden-Rossendorf, Germany
\aff{2}Department of Earth, Planetary and Space Sciences, University of California, Los Angeles,
CA 90095, USA
\aff{3}Centre for Fluid and Complex Systems, Coventry University, Coventry CV1 5FB, United Kingdom}

\begin{document}
\maketitle

\begin{abstract}
We present the first detailed thermal and velocity field characterization of convection in a rotating cylindrical tank of liquid gallium, which has thermophysical properties similar to those of planetary core fluids. Our laboratory experiments, and a closely associated direct numerical simulation, are all carried out in the regime prior to the onset of steady convective modes. This allows us to study the oscillatory convective modes, sidewall modes and broadband turbulent flow that develop in liquid metals before the advent of steady columnar modes.  Our thermo-velocimetric measurements show that strongly \JA{inertial, thermal wind} flows develop, with velocities reaching those of comparable non-rotating cases. Oscillatory bulk convection and wall modes coexist across a wide range of our experiments, along with strong zonal flows that peak in the Stewartson layer, but that extend deep into the fluid bulk in the higher supercriticality cases. The flows contain significant time-mean helicity that is anti-symmetric across the midplane, demonstrating that oscillatory liquid metal convection contains the kinematic components to sustain system-scale dynamo generation. 
\end{abstract}
 
\section{Introduction}
The geomagnetic field is induced by the liquid metal flow within Earth's outer core via self-excited dynamo action. Thermal and compositional buoyancy drives the fluid motion because the iron-rich core is cooling from its primordial state through heat loss to the mantle \citep{Jacobs1953, Davies2015}. The detailed flow topology is unknown \citep{Calkins2012, Guervilly2016, Aurnou2017, Kaplan2017, Guervilly2019}, since the \SI{3000}{km} thick silicate mantle hinders our ability to observe core dynamics directly \citep{Roberts2013}. Furthermore, thermally-driven global scale dynamo models in low Prandtl number fluids, characteristic of liquid metals, can not be carried out by direct numerical simulations (DNS) to date \citep{Roberts2013, Davies2009, Nataf2015, Schaeffer2017}. 

Our understanding of the flow dynamics in Earth's outer core must instead rely on theory, experiments and numerical simulations under simplified conditions. Towards this end, we investigate the effect of rotation on low Prandtl number thermal convection by means of laboratory experiments and DNS. We consider a rotating Rayleigh-B\'{e}nard convection set-up, consisting of a cylindrical vessel filled with liquid metal where the convective flow is driven by a temperature difference between a warmer bottom and a colder top boundary. The cylinder rotates around its vertical axis, which is aligned parallel with gravity, roughly similar to the high latitude regions within Earth's outer core \citep[i.e.,][]{Cao2018}. 

Rotating Rayleigh-B\'enard convection is controlled by three dimensionless parameters. The Rayleigh number $Ra$ describes the ratio of buoyancy to thermal and viscous diffusion and is defined as:
 \begin{equation}
Ra  =  \alpha g \Delta T H^3 / (\kappa \nu) \, ,  
\label{eq:Ra}
\end{equation}
where $\alpha$ denotes the thermal expansion coefficient, $g$ is the gravitational acceleration, $\Delta T$ is the applied temperature difference between the bottom and top of the fluid layer, $H$ is the height of the fluid layer, $\kappa$ is the thermal diffusivity and $\nu$ is the kinematic viscosity. In rotating systems, the Ekman number $Ek$ describes the ratio of the viscous and Coriolis forces:
 \begin{equation}
Ek  =  \nu / (2 \Omega H^2) \, ,
\label{eq:Ek}
\end{equation}
where $\Omega$ is the system's angular rotation rate. Whereas $Ra$ and $Ek$ estimate the forces involved, the Prandtl number $Pr$ describes the characteristics of the fluid, and is given by the ratio of the viscous and thermal diffusivities:
 \begin{equation}
Pr  =  \nu / \kappa \, .
\label{eq:Pr}
\end{equation}

The majority of experiments and numerical simulations of rotating convection consider fluids with Prandtl number values of \TV{${\mathcal{O}(1)}$}, corresponding to fluids like air ($Pr \simeq 0.7$) or water ($1.75 \leq Pr \leq 13.5$) \citep{Zhong1993, Julien1996, King2012, Kunnen2010, Stevens2010, Zhong2010, Weiss2011a, Stevens2013, Ecke2014, Horn2014, Cheng2015}. In the laboratory, these fluids are typically easy to handle experimentally, and are also accessible for optical velocimetric measurements \citep[e.g.,][]{Kunnen2010, Aujogue2018}. Studying convection in $Pr \approx 1$ fluids is attractive from a numerical point of view because the numerical costs are much lower than for high and low $Pr$ fluids at the same boundary conditions \citep{Shishkina2010, Calkins2012, Horn2017}. For this reason, the majority of present-day rotating convection and dynamo simulations employ $Pr \approx 1$ fluids, and, hence disregard possible \TV{$Pr \ll 1$} effects \cite[cf.~][]{Aubert2001, King2013, Kaplan2017, Bouffard2019, Guervilly2019}.

The critical Rayleigh number, $Ra_c$, at which convection first develops, or onsets, generally increases with $\Omega$, since rotational effects typically inhibit convection in rotating systems \citep[][cf.~{Horn2018, Horn2019}]{Julien1998, Zhang2000, Chandrasekhar1961}. In moderate $Pr$ fluids, convection onsets via flow modes that have their highest amplitude close to the sidewall, and are thus called wall modes \citep{Zhong1991, Ecke1992, Ning1993, Zhong1993, Liu1997, Liu1999}. At moderate supercriticalities, $Ra / Ra_c$, the bulk flow is dominated by elongated columnar structures, that are called convective Taylor columns or Ekman vortices \citep{Nakagawa1955, Julien1996, Sakai1997, Sprague2006, Kunnen2008, Stevens2009, Grooms2010, Weiss2010, Zhong2010, King2012, Stevens2013, Cheng2015, Gastine2016, Guervilly2019}. These quasi-steady columnar structures are the key magnetically inductive flow component in present day dynamo simulations \citep{Olson1999, Roberts2013, Aurnou2015, Christensen2015}.

The Prandtl number in a liquid metal is typically $Pr \approx O(10^{-2})$, indicating that temperature diffuses much faster than momentum. As a consequence, the inertia dominated velocity field in liquid metal convection tends to become turbulent at moderate supercriticalities, while the temperature field is characterized by larger, and more coherent patterns \citep[e.g][]{Vogt2018jrv}. In low $Pr$ fluids, however, rotating convection is inherently different and far less well understood. For $Pr \ll 1$ rotating convection, the bulk onset mode is oscillatory \citep{Zhang2009, Horn2017, Aurnou2018}. It is often presumed that since oscillatory modes are inefficient transporters of heat, they can only generate weak convective motions that will be easily overwhelmed by the stationary modes that are excited at larger supercriticalities \citep[e.g.][]{Roberts2013}. Based on such arguments, low $Pr$ oscillatory convective flows are often ignored in models of planetary dynamo action.

Prior to the advent of dynamo simulations, however, it was argued that small-scale inertial oscillations could drive global scale dynamo action \citep[e.g.][]{Moffatt1978, Olson1977, Olson1983}, and more recently in \citet{Calkins2015} and \citet{Davidson2018}. Rapidly rotating, kinematic plane layer dynamo models have now demonstrated that low $Pr$ oscillatory convective modes are actually capable of generating dynamo action at lower $Ra$ and in lower electrical conductivity fluids than in moderate $Pr$ cases \citep{Calkins2016a, Calkins2016b}.

The aim of this work is to shed further light on the fluid dynamics of low $Pr$ rotating convection via combined laboratory-numerical experiments. We follow up on our previous experimental investigation, \cite{Aurnou2018}, in which low $Pr$ rotating convection was investigated by means of point-wise temperature measurements in the bulk and at the sidewall of a cylindrical convection vessel. The thermal measurements indicated that convection sets in first via oscillatory modes, in good agreement with theory \citep{Zhang2009}. At slightly higher supercriticalities, wall modes were detected that coexisted with the oscillatory bulk modes.
Furthermore, broad band turbulence was inferred to develop well below the onset of steady convection modes. Hence, the quasi-steady convective Taylor columns that dominate rotating convection at $Pr \gtrsim 1$ were not found to dominate in these $Pr \ll 1$ experiments, in agreement with \citet{Horn2017}.

Considering the potential importance of this finding with respect to the flow structures that underlie planetary dynamo generation, here we extend our low $Pr$ rotating convection experimental system to include velocity measurements by means of ultrasound Doppler velocimetry (UDV) and complement our laboratory experiments with direct numerical simulation results. We investigate a parameter range comparable to the thermometric experiments of \citet{Aurnou2018}. \JA{Our UDV and DNS results demonstrate that the thermal-inertial oscillatory convection velocities simultaneously attain rotationally-dominated thermal wind values, $u_{TW} = \alpha g \Delta T/(2 \Omega)$, and near free-fall values, $u_{f\!f}=\sqrt{\alpha g \Delta T H}$, for which thermal buoyancy is transferred completely into fluid inertia \citep[cf.][]{Aurnou2020}.} Multi-modal bulk oscillations dominate the velocity field over the whole range of supercriticalities investigated. Additionally, coherent time-mean zonal flows and time-mean helicity is found in the rotating liquid metal convection, showing that the essential ingredients for dynamo generation \citep{Roberts2015} are already present in oscillatory, low $Pr$ rotating convection.


\section{Onset predictions for low \textit{Pr} rotating convection \label{sec:onset}}

In rotating systems, thermal buoyancy has to overcome the stabilizing effect of the Coriolis force in order to inititate convective flow instabilities. The critical Rayleigh number $Ra_c$ for the onset of convection increases with decreasing $Ek$ \citep{Chandrasekhar1961}.  
\SH{
Convective instability sets in first via bulk thermal-inertial oscillations ---not wall modes and not steady columns--- in fluids with $Pr \leq 0.68$ and $Ek \gtrsim 10^{-7}$ \citep{Horn2017, Aurnou2018}.} 
These oscillations can be described by a balance between the inertial, Coriolis and pressure gradient forces \citep{Zhang2017}. In a horizontally infinite plane ($\infty$), the oscillatory ($O$) convection is predicted to first develop, or onset, at:
\begin{equation}
Ra^{\infty}_{O}  
\simeq 17.4 \, \left( Ek/Pr \right)^{-4/3} \, .
\label{eq:RaO}
\end{equation}
\citep{Chandrasekhar1955, Chandrasekhar1961, Julien1998, Zhang1997}. The oscillation frequency at the onset of this oscillatory convection is:
\begin{equation}
\label{eq:fo}
\widetilde{f}^{\infty}_{O} = {f}^{\infty}_{O} / f_\Omega 
\simeq 4.7 \left( Ek / Pr \right)^{1/3} \, .
\end{equation}
Frequencies marked with a tilde are normalized by the characteristic rotation frequency $f_\Omega = 1/T_\Omega=\Omega/(2 \pi)$ throughout this work, i.e. $\widetilde{f} \equiv f/f_\Omega$.
The horizontal length scale of the oscillatory mode, measured perpendicular to the rotation direction $\hat{z}$, is given by:
\begin{equation}
\label{eq:length}
\ell_O^\infty 
\simeq 2.4 \,  \left( Ek / Pr \right)^{1/3} H \, .
\end{equation} 
There exist asymptotic predictions for finite cylindrical ($cyl$) fluid volumes at low $Pr$ and $Ek$ from \cite{Zhang2009}. Their equations (4.21) and (4.22) yield more accurate estimates for the onset Rayleigh number $Ra^{cyl}_{O}$ and the oscillation frequency $\widetilde{f}^{cyl}_{O}$. Both values approach the values estimated with (\ref{eq:RaO}) and (\ref{eq:fo}) as $Ek \rightarrow 0$, but far more slowly than in $Pr>1$ fluids \citep{Goldstein1994}. In this study, we define convective supercriticality as
 \begin{equation}
\label{eq:super}
 \widetilde{Ra}=Ra / Ra^{cyl}_{O},  
\end{equation}
such that  $\widetilde{Ra} > 1$ for convection to onset. 

The sidewalls of a cylindrical container can help to overcome the stabilizing effect of rotation, leading to the formation of wall modes \citep{Ecke1992, Herrmann1993, Kuo1993, Zhang2017}. The onset of convective wall modes is predicted to occur at \citep{Zhang2009}:
\begin{equation}
Ra_W  
\simeq 31.8 Ek^{-1} + 46.6 Ek^{-2/3}
\label{eq:RaW}
\end{equation}
and they travel in the retrograde azimuthal direction with a drift frequency of:
\begin{equation}
\label{eq:fw}
\widetilde{f}_{W} 
\simeq 132.1 (Ek / Pr) - 1465.5(Ek^{4/3} / Pr) \, .
\end{equation}
%
\SH{The higher-order correction terms in \eqref{eq:RaW} and \eqref{eq:fw} were derived by \citet{Zhang2009} to account for no-slip boundary conditions, building on the free-slip formulations of \citet{Herrmann1993}. Both studies assume a semi-infinite domain \JA{such that there is zero  curvature of the sidewall.} Both sets of predictions are shown to be accurate for $Ek \lesssim 10^{-3}$ in the current cylindrical set-up with $\Gamma = 1.87$ \citep{Horn2017}. \citet{Horn2015} also showed good agreement with these predictions in a $Pr = 0.8$ set-up with $\Gamma = 0.5$. For even smaller aspect ratios, however, curvature effects become important, leading to notably higher $Ra_W$ values \JA{(unpublished DNS)}.}

Since wall modes form on the thermally passive sidewalls of cylindrical containers, they do not have an exact equivalent in spherical geometries \citep[e.g.,][]{Aurnou2018}. However, in a recent experimental study it has been argued that slowly oscillatory, wall-mode-like flows form at the virtual boundary of the tangent cylinder, which is the virtual cylinder circumscribing Earth's solid inner core and aligned with the axis of rotation \citep{Aujogue2018}.

The onset of steady convection is predicted, for $Ek \ll 1$, when the Rayleigh number exceeds the critical value
\begin{equation}
\label{eq:Ras}
Ra^{\infty}_{S} 
\simeq 8.7 Ek^{-4/3}, 
\end{equation}
with horizontal length scale
\begin{equation}
\label{eq:ls}
\ell^{\infty}_{S} 
\simeq 2.4 Ek^{1/3} \, H
\end{equation}
%
\citep{Chandrasekhar1961, Julien1998}. 

A comparison of the onset conditions of the different low $Pr$ rotating convective modes show that the bulk oscillatory convective instability sets in before the wall modes and well before the onset of stationary convection in our experiments. Here, $Ra^{\infty}_{S}/Ra_O^{cyl} \approx 20$, whereas our highest supercriticality experimental case occurs at $\widetilde{Ra} = Ra/Ra_O^{cyl} = 15.7$.  Thus, we investigate the properties of bulk oscillatory convection and drifting sidewall convective modes, without the effects of the stationary convection modes that typically dominate higher $Pr$ rotating convection systems.

\section{Laboratory-numerical system \label{sec:lns}}
\subsection{Laboratory set-up}\label{sec:setup}
\begin{figure}
  \centerline{\includegraphics[width=1\textwidth]{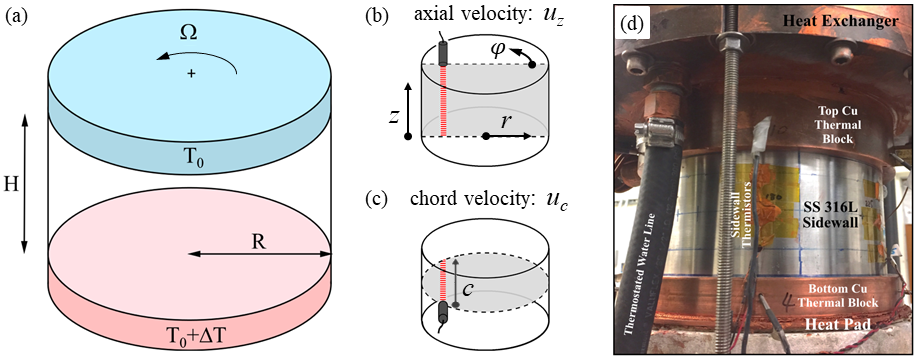}}
  \caption{Schematic of the experimental set-up (\textit{a}) and the UDV Sensor positions for (\textit{b}) the axial velocity and (\textit{c}) the chord velocity measurements. (\textit{d}) Photograph of the experimental set-up without sidewall insulation in place. Image credit: Yufan Xu (UCLA).}
\label{fig:setup}
\end{figure}

Figure \ref{fig:setup} (\textit{a}) shows a schematic drawing of the experimental set-up. The experiments were performed in a cylindrical vessel having an inner height of $\textit{H}=98.4$ mm and inner radius of $\textit{R}=98.4$ mm which gives an aspect ratio of $\Gamma=2R/H=2$. The sidewall is made of stainless steel and several layers of thermal insulation. The bottom plate is made of copper that is heated from a non-inductively wound electrical heatpad, whereby the input power ranges from $P=50 W$ to $800 \mathrm{W}$ in this study. The heat is removed from the top copper plate by a circulating water bath. The whole set-up is mounted on a turntable that allows for a rotation around the upright axis of the cylindrical vessel. We investigated four rotation rates: $4.08$ rpm, $8.16$ rpm, $16.33$ rpm and $32.66$ rpm, corresponding to $Ek = 4 \times 10^{-5}, 2 \times 10^{-5}, 1 \times 10^{-5}$ and $5 \times 10^{-6}$, respectively. The experimental set-up is similar to that described in \cite{Aurnou2018}, except for their use of an acrylic sidewall material. A detailed description of the device can be found in \cite{King2012}.

The container is filled with the liquid metal gallium. The material properties of gallium were adopted from \cite{Aurnou2018}. It has a melting temperature of $T_{mp}=29.8 \, \mathrm{^\circ}$C, with a melting point density of $\rho_{mp}=6.09 \, \times10^3  \, \mathrm{kg/m^3}$. The thermal conductivity is $k=31.4 \,  \mathrm{W/(mK)}$, the thermal expansion coefficient is $\alpha = 1.25\times10^{-4}  \, \mathrm{K^{-1}}$ and the specific heat capacity is $c_p=397.6  \, \mathrm{J/(kg K)}$. The dynamic viscosity of gallium is described by $\mu=\mu_0 \, \mathrm{exp}\left(E_a / RT\right)$, whereby $\mu_0=0.46  \, \mathrm{mPa \, s}$ is the viscosity coefficient, $E_a = 4000  \, \mathrm{J/mol}$ is the activation energy, and $R=8.3144 \,  \mathrm{J/(mol \, K)}$ is the gas constant. Typical kinematic viscosity and thermal diffusivity values in our experiments are $\nu = \mu / \rho = 3.4 \times 10^{-7}$ $\mathrm{m^2/s}$ and $\kappa = k/(\rho \, c_p) = 1.3 \times 10^{-5}$ $\mathrm{m^2/s}$, corresponding to a characteristic Prandtl number value of $Pr \simeq 0.026$.

\subsection{Measuring technique}
\label{tech}
\begin{center}
\begin{figure}
 \includegraphics[width=1\textwidth]{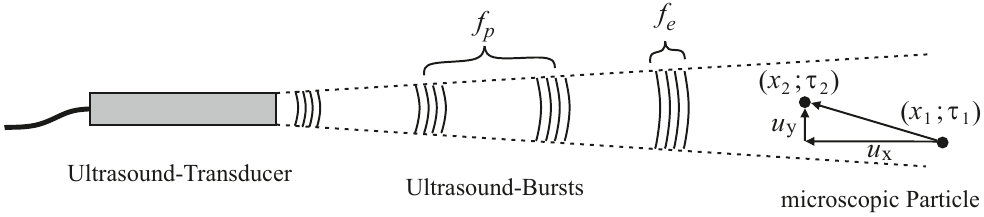}
 \caption{Ultrasonic Doppler Velocimetry schematic. The pulse repetition frequency is $f_p$, the frequency of the ultrasound burst is $f_e$, and $u_x$ and $u_y$ are the velocities parallel and perpendicular to the ultrasonic beam, respectively. The distance from the ultrasound transducer is $x_i$ at measurement time $\tau_i$.}
 \label{fig:UDV}
\end{figure}
\end{center}

The main experimental results in this study are provided by spatio-temporal velocity recordings based on ultrasonic Doppler velocimetry (UDV). The UDV technique uses short ultrasound bursts that are transmitted from a piezo-electric transducer into the liquid metal (Figure \ref{fig:UDV}).  After each burst, the transducer acts as a receiver that detects echoes emitted from microscopic impurities contained in the liquid metal. These impurities are primarily oxides that exist in non-precious liquid metals like gallium and do not need to be added. The pulse emission and the subsequent echo recording are repeated periodically. This allows one to calculate the distance $x$ between particles and transducer based on the time delay $\tau$ between an emitted burst and its echo
\begin{equation}
	x= c_s \, \tau / 2 \, ,
\end{equation}
where $c_s$ is the fluid's speed of sound. If the particle moves with the fluid between two bursts, the position change is related to the velocity as follows
\begin{equation}
	x_2 - x_1 = c_s \, (\tau_2 - \tau_1) / 2  = u_x / f_p 
\label{udv2}
\end{equation}
where $f_p$ is the pulse repetition frequency of the ultrasonic wavepackets. Since the time difference $(\tau_2 - \tau_1)$ is typically rather small, the measuring system instead utilizes the more easily measured phaseshift
\begin{equation}
	\delta = 2 \pi f_e (\tau_2 - \tau_1)
	\label{udv3}
\end{equation}
where $f_e$ is the emitted ultrasound frequency. From (\ref{udv2}) and (\ref{udv3}), the particles beam-parallel velocity is calculated as
\begin{equation}
	u_x = c_s \, \delta f_p / (4 \pi f_e) \, .
\end{equation}

\begin{figure}
  \centerline{\includegraphics[width=0.9\textwidth]{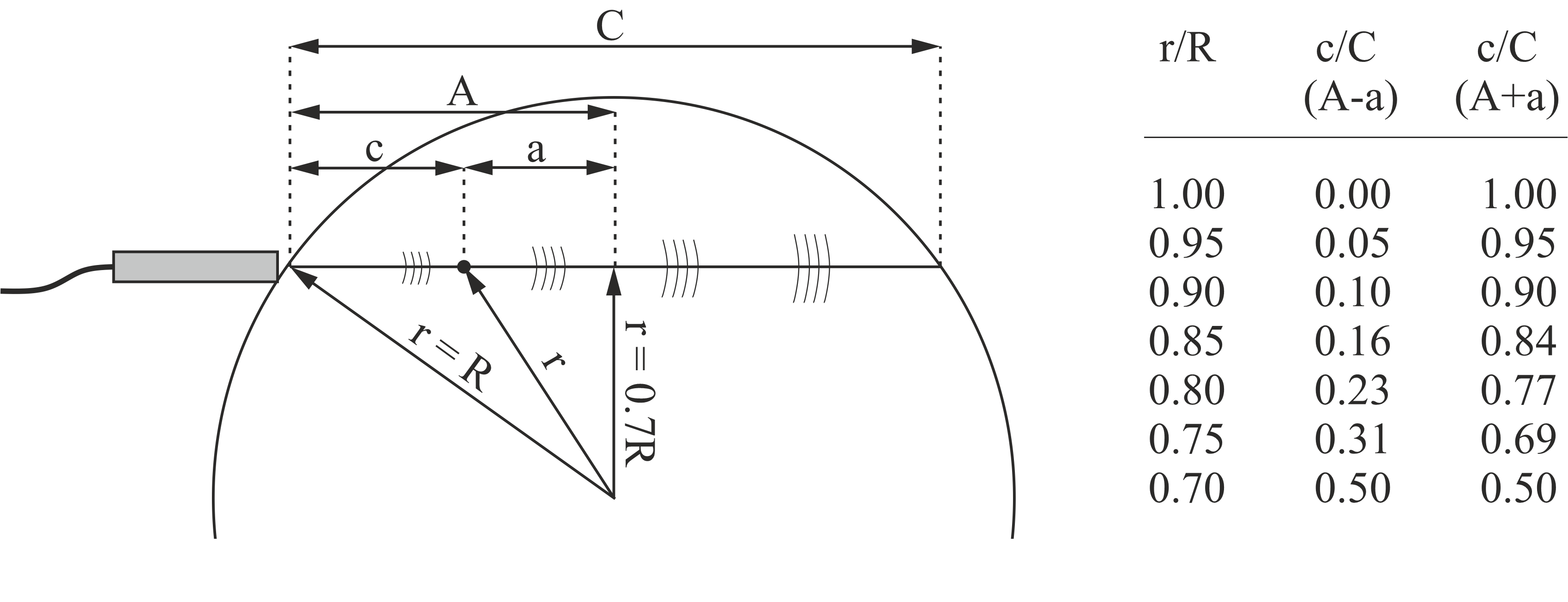}}
  \caption{\TV{Schematic representation of expression (\ref{chord_equation}), which relates the distance along chord-probe beam $c$ and the radial coordinate $r$.}
\label{fig:Chord_schematic}}
\end{figure}

Ultrasonic Doppler velocimetry is a well-established tool to measure velocity profiles in opaque fluids such as liquid metals \citep[see][]{Brito2001, Aubert2001, Eckert2002, Gillet2007, Nataf2008, Vogt2013spin, Vogt2014inertial, Tasaka2016, Vogt2018jrv, Vogt2018transition}. The UDV system employed in this study is the DOP3010 (Signal Processing SA, Lausanne), equipped with $f_e = 8$ MHz piezo-electric transducers. One transducer is located in the top copper plate at $r/R=0.7$ and measures the vertical velocity (Figure \ref{fig:setup}\textit{b}). A second transducer is located halfway up the vessel and measures the velocity along a chord that crosses the beam of the vertical transducer in the midplane (Figure \ref{fig:Chord_schematic}). 

The chord is related to the radius as follows: 
\TV{\begin{equation}
	\frac{c}{C} = \frac{A \pm a}{2A} = \frac{\sqrt{R^2-(0.7 R)^2} \pm \sqrt{r^2-(0.7 R)^2}}{2\sqrt{R^2-(0.7 R)^2}}.
	\label{chord_equation}
\end{equation}}
The measuring system records velocity profiles with a temporal resolution of about $0.3$ s. The spatial resolution is about $1$ mm in the beam direction and about $5$ mm in the lateral direction due to the diameter of the ultrasound emitting piezo.

The experiment is also equipped with a total of 30 thermistors with six thermistors embedded in the top plate and six embedded in the bottom plate, which measure the temperature drop across the fluid layer. One thermistor is immersed in the liquid metal bulk, whose data are used in temperature FFTs. The remaining seventeen thermistors are attached on the outside of the stainless steel \JA{sidewall, similar to \citet{Aurnou2018}.} All temperature data is acquired at a 10 Hz sampling rate.

 \subsection{Direct numerical simulations}
Direct numerical simulations (DNS) are performed using the fourth-order accurate finite volume code \textsc{goldfish} \citep{Shishkina2015, Horn2017}. \SH{In} the DNS, the non-dimensional, incompressible Navier-Stokes equation are numerically solved together with the temperature equation in the Oberbeck--Boussinesq approximation:
\begin{subequations}
\label{eq:NavierStokes}
\begin{eqnarray}
 \nabla \cdot {\vec{u}} &=& 0,\\
D_t \vec{u} &=&  \sqrt{\frac{Pr}{Ra}} \vec{\nabla}^2 {\vec{u}}  - \vec{\nabla} {p} - \sqrt{\frac{Pr}{Ra\, Ek^2}} {\hat{\vec{e}}_z} \times {\vec{u}} + {T} {\hat{\vec{e}}}_z, \\
D_{{t}} {T} &=&   \sqrt{\frac{1}{Ra\, Pr}}  \vec{\nabla}^2 {T}.
\end{eqnarray}
\end{subequations}
The reference scales for the DNS non-dimensionalisation are the temperature difference $\Delta T$, the fluid layer height $H$, and the free-fall velocity $u_{f\!f} = \sqrt{\alpha g \Delta T H}$. The top and bottom boundaries are perfectly isothermal and the sidewall is thermally insulating; no-slip, impenetrable velocity boundary conditions are enforced on all walls. 
\SH{The focus of the numerical analysis is on a single canonical case with $Pr = 0.025$, $\Gamma = 1.87$, $Ra = 8 \times 10^6$ and $Ek = 5\times 10^{-6}$, corresponding to $\widetilde{Ra} = 2.30$. Complementary analyses of this DNS have been published previously \citep{Horn2017}. Here, we continued this simulation 
to save snapshots at a finer sampling rate of 10 snapshots per free-fall time unit to produce synthetic Dopplergrams with a comparable temporal resolution as the experiments. All data analysed in this paper originate from this new data set. The slight mismatch of $\Gamma$ between DNS and laboratory experiments appears to have only a minor effect and still allows for good quantitative comparison with laboratory experiments \citep[e.g.][]{Vogt2018jrv}}.

 \section{Results \label{sec:results}}
 \subsection{Global heat and momentum transport}

\begin{figure}
\includegraphics[width=\textwidth]{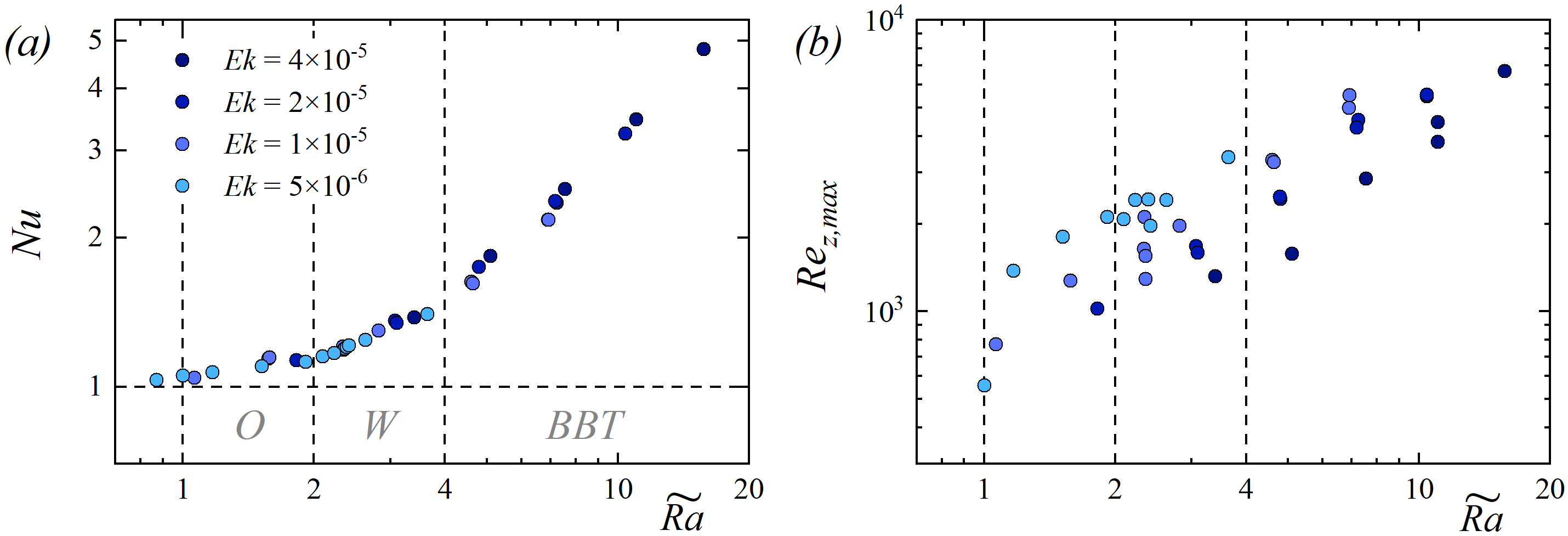}
\caption{(a) Nusselt number, $Nu$, plotted as a function of convective supercriticality $\widetilde{Ra}$. The corresponding Ekman numbers are indicated with the symbol color. The abbreviation $O$ denotes oscillatory bulk, $W$ denotes wall modes and $BBT$ stands for broad band turbulence. (b) Reynolds number $Re_{z,max}$ based on the vertical velocity maximum versus $\widetilde{Ra}$.}
\label{fig4}
\end{figure}

Laboratory experiments are made at four different $Ek$, with the highest $\widetilde{Ra}$ achieved in the highest $Ek$ cases. All data are recorded during the equilibrated state in which the mean fluid temperature is constant. Figure \ref{fig4}(\textit{a}) shows the measured convective heat transfer, expressed non-dimensionally in terms of the Nusselt number, which is the ratio of the total vertical heat flux $Q_{tot} = P / (\pi R^2)$ and the purely conductive heat flux $Q_{cond} = k \Delta T/H$:
\begin{equation}
Nu = PH/ (\pi k \Delta T R^2) , 
	\label{nu}
\end{equation}
plotted as a function of supercriticality $\widetilde{Ra}$. It was shown by \citet{Aurnou2018} that experiments at different Ekman numbers can be best compared if they are plotted versus the supercriticality $\widetilde{Ra}$. This is confirmed by our experiments as shown in figure \ref{fig4}$(a)$. Although we used less sidewall insulation in this suite of experiments, good agreement is found between our present $Nu$ data and that of \citet{Aurnou2018}.

Four different states can be identified in the range of $\widetilde{Ra}$ considered. At $\widetilde{Ra}\leq 1$ the system is subcritical and the heat transfer is purely conductive. Convection sets in at $\widetilde{Ra} \simeq 1$ in the form of thermal-inertial oscillations in the fluid bulk. Wall-modes develop at $\widetilde{Ra}\simeq 2$, leading to an increased heat transport scaling efficiency. Above $\widetilde{Ra}\simeq 4$, the determination of individual modes becomes difficult since the flow and temperature fields are increasingly determined by broadband thermal and velocity signals. The onset of steady bulk convection is predicted at $\widetilde{Ra} = Ra_{S}/Ra^{cyl}_{O} \approx 20$, which is beyond the range of supercriticalities investigated in this study.

Figure \ref{fig4}(b) shows UDV measurements in liquid metal rotating convection. 
Figure \ref{fig4}(b) presents Reynolds numbers, $Re_{z,max} = u_{z,max} \, H / \nu$, calculated using the peak velocity depth-averaged over the range $0.45 < z/H < 0.55$ on the vertical ultrasonic transducer  located at a radial position $r/R \approx 2/3$ (see figure \ref{fig:setup}(b)). These raw $Re_{z,max}$ data are not well collapsed, and instead are separated by their $Ek$-values \TV{with light blue, low $Ek$ values at the top of the data and dark blue, high $Ek$ values at the bottom of the data.} However, for all the data beyond $\widetilde{Ra} \simeq 1.2$, we find $Re_{z,max} > 10^3$. Thus, rather intense flows develop just after the onset of convection, as characteristic of low $Pr$ fluids \citep{Clever1981, Grossmann2008, Calkins2015, Vogt2018jrv}. 

 \subsection{Velocity scalings}
\begin{figure}
\includegraphics[width=\textwidth]{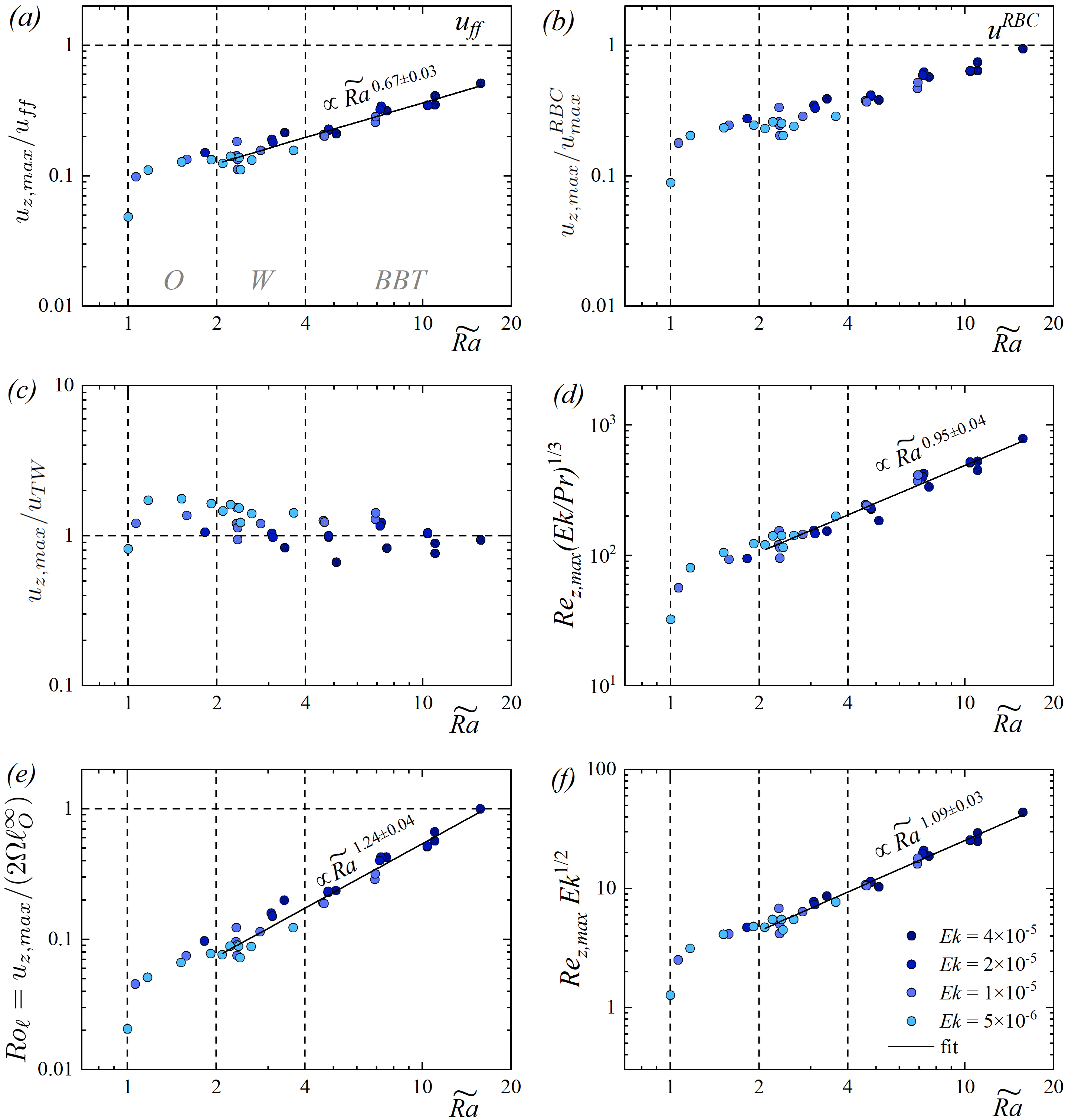}
\caption{\JA{Normalized maximum vertical UDV velocities, $u_{z, max}$, plotted versus convective supercriticality $\widetilde{Ra}$. (a) $u_{z, max}$ normalized by the inertial free-fall velocity $u_{f\!f} = \sqrt{\alpha g \Delta T H}$. (b) $u_{z, max}$ normalized by the maximum velocity scaling in the non-rotating liquid gallium RBC experiments of \citet{Vogt2018jrv}. (c) $u_{z, max}$ normalized by the thermal wind velocity $u_{TW} = \alpha g \Delta T/ (2 \Omega)$. (d) Local Reynolds number based on $u_{z, max}$ and the bulk oscillatory length scale estimate $\ell_O^\infty \sim (E/Pr)^{1/3} H$. (e) Local Rossby number based on $u_{z, max}$ and $\ell_O^\infty$ given by \eqref{eq:length}. (f) Local Reynolds number based on $u_{z, max}$ and the approximate Ekman layer thickness $\ell_{Ek} \sim Ek^{1/2} H$.}}
\label{f:scalings}
\end{figure}

Figure \ref{f:scalings}(a) plots the vertical UDV data normalized with \JA{the free-fall velocity $u_{f\!f} = \sqrt{\alpha g \Delta T H}$, which is the upper bounding velocity in an inertially-dominated convection system \citep[e.g.,][]{King2013}.} The first measurable flow appeared at $\widetilde{Ra} = 1.002$ and reaches velocity values of $u_{z,max}/u_{f\!f} \simeq 0.1$ for $\widetilde{Ra} > 1$. \JA{The data suggests a $u_{z,max}/u_{f\!f} \propto \widetilde{Ra}^{2/3}$ power-law trend for $\widetilde{Ra} > 2$.} Interestingly, the vertical velocities already reach 50$\%$ of the free-fall velocity estimate, $u_{z,max}/u_{f\!f} = 0.51$, in our moderately supercritical $\widetilde{Ra} = 15.7$ case. 
\JA{This demonstrates that strongly inertial flows develop in moderate $\widetilde{Ra}$, low $Pr$ rotating convection experiments.}


Figure \ref{f:scalings}(b) shows the $u_{z,max}$ data normalized by the maximum velocity scaling, $Re_{max}^{RBC} = 0.99 (Ra/Pr)^{0.483}$, for the non-rotating $Pr \approx 0.026$ Rayleigh-B\'enard convection experiments carried out by \citet{Vogt2018jrv} in the same experimental set-up. \JA{The highest supercriticality case attains nearly the same maximum velocity as found in non-rotating RBC at the same $Ra$, $u_{z,max} = 0.94 \, u_{max}^{RBC}$.} This case also corresponds to a local Rossby number $Ro_\ell = u_{z,max}/(2 \Omega l_O^\infty)  \simeq 1$, where the local-scale flow should be only marginally constrained by rotational effects. At this time, it remains unknown whether, in higher supercriticality cases, the oscillatory convection velocities can exceed the near free-fall low $Pr$ RBC velocities, or if the velocity scaling will flatten out such that $u_{z,max}  \simeq u_{max}^{RBC}$ in the $Ro_\ell \gtrsim 1$ regime. 

\JA{Figure \ref{f:scalings}(c) plots $u_{z,max}$ normalized by the thermal wind velocity, $u_{TW} = \alpha g \Delta T / (2 \Omega)$, which should be the dominant flow velocity when the dynamics are strongly controlled by the system's rotation.  With this normalization, all the data are clustered in the vicinity of unity ($0.7 < u_{z,max}/u_{ff} < 1.8$), showing that the thermal wind scaling holds well. Comparing figures \ref{f:scalings}(a-c) shows that our thermal-inertial flows are simultaneously in thermal wind balance and in inertial balance.  Thus, these $Ra < Ra_S^\infty$, low $Pr$ flows are well described by a Coriolis-inertial-Archimedian (CIA) triple balance, as is argued to be relevant in planetary core bulk dynamics \citep[e.g.,][]{Aubert2001, Christensen2006, Jones2011, KingBuffett2013, Gastine2016, Long2020}. 
}

%
\JA{
CIA dynamics imply that the UDV data should collapse using a thermal-wind based, local Reynolds number, since the lateral width of the bulk convective modes is the dynamically relevant scale in rapidly rotating vorticity dynamics \citep[e.g.,][]{Calkins2018}. This local Reynolds number should, in turn, scale linearly with the convective supercriticality \citep{Maffei2020}: 
\begin{equation}
Re_\ell = \frac{u_{TW} \ell_O^\infty}{\nu} \sim  \frac{Ra \, Ek^{4/3}}{Pr^{4/3}}  \sim \widetilde{Ra}\, . 
\label{Re_ell}
\end{equation}
Figure \ref{f:scalings}(d) tests \eqref{Re_ell} by plotting $Re_{z,max} (Ek/Pr)^{1/3}$ versus $\widetilde{Ra}$. The best fit to the $\widetilde{Ra} > 2$ data is in good agreement with \eqref{Re_ell}'s linear scaling prediction, in further support that our flows exist in a thermal-inertial, CIA-style balance. For ease of cross-comparison, expression \eqref{Re_ell} can be converted to its system scale counterpart, yielding
\begin{equation}
Re = Re_\ell \, \frac{H}{\ell_O^\infty}  \sim \frac{Ra \, Ek}{Pr} \, ,
\label{E:Re}
\end{equation}
in agreement with \citet{Guervilly2019}, \citet{Maffei2020}, and \citet{Aurnou2020}.
}

\JA{In contrast to figure \ref{f:scalings}(d), the $u_{z,max}$ measurements are not well collapsed by the system-scale Rossby number, $Ro = u_{z,max} / (2 \Omega H)$, which estimates the ratio of flow inertia and system-scale Coriolis force. In our experiments, the Rossby number values lie in the range $ 3 \times 10^{-3} \lesssim Ro \lesssim 3 \times 10^{-1}$.  The $Ro$ data are spread out nearly as strongly as the $Re_{z,max}$ data in figure \ref{fig4}(b).  In figure \ref{f:scalings}(e), we instead plot the local Rossby number, $Ro_\ell = u_{z,max}/(2 \Omega \ell_O^\infty)$.  The UDV data are moderately well collapsed by $Ro_\ell \propto \widetilde{Ra}^{5/4}$. This best fit can be explained by noting that $Ro_\ell \sim Ra (Ek/Pr)^{5/3}$, whereas $\widetilde{Ra}^{5/4} \sim Ra^{5/4} (Ek/Pr)^{5/3}$.  Thus, $Ro_\ell$ and $\widetilde{Ra}^{5/4}$ have identical $Ek$ and $Pr$ scalings and differ only by a factor of $Ra^{1/4}$. (Similarly structured arguments can be used to explain the $\widetilde{Ra}^{2/3}$ scaling in figure \ref{f:scalings}a.)}

\JA{We note further that the figure \ref{f:scalings}(e) data lie in the $0.1 < Ro_\ell \leq 1$ range.  In this $Ro_\ell = \mathcal{O}(1)$ regime, the thermal wind and free fall velocities should be comparable \citep{Aurnou2020}, which explains why these low $Pr$ rotating convection velocities are approaching the non-rotating free-fall limit.}
 
\JA{Figure \ref{f:scalings}(f) plots the local Reynolds number based on the Ekman boundary layer thickness, $\lambda_{Ek} \simeq Ek^{1/2} H$. This yields $Re_{z,max} \, \lambda_{Ek} /H = Re_{z,max} Ek^{1/2}$.  Naively, one might speculate that the thermal-inertial data would be insensitive to viscous boundary layer processes.  However, the data in figure \ref{f:scalings}(f) are fairly well collapsed by $Re_{z,max} Ek^{1/2}$.  This suggests, somewhat nonintuitively, that the effects of Ekman boundary layers and viscous dissipation are not negligible in these low $Pr$ rotating convection experiments \citep[cf.][]{Stellmach2014, Julien2016, Plumley2016, Plumley2017, Maffei2020}.
}

 \begin{figure}
 \includegraphics[width=\textwidth]{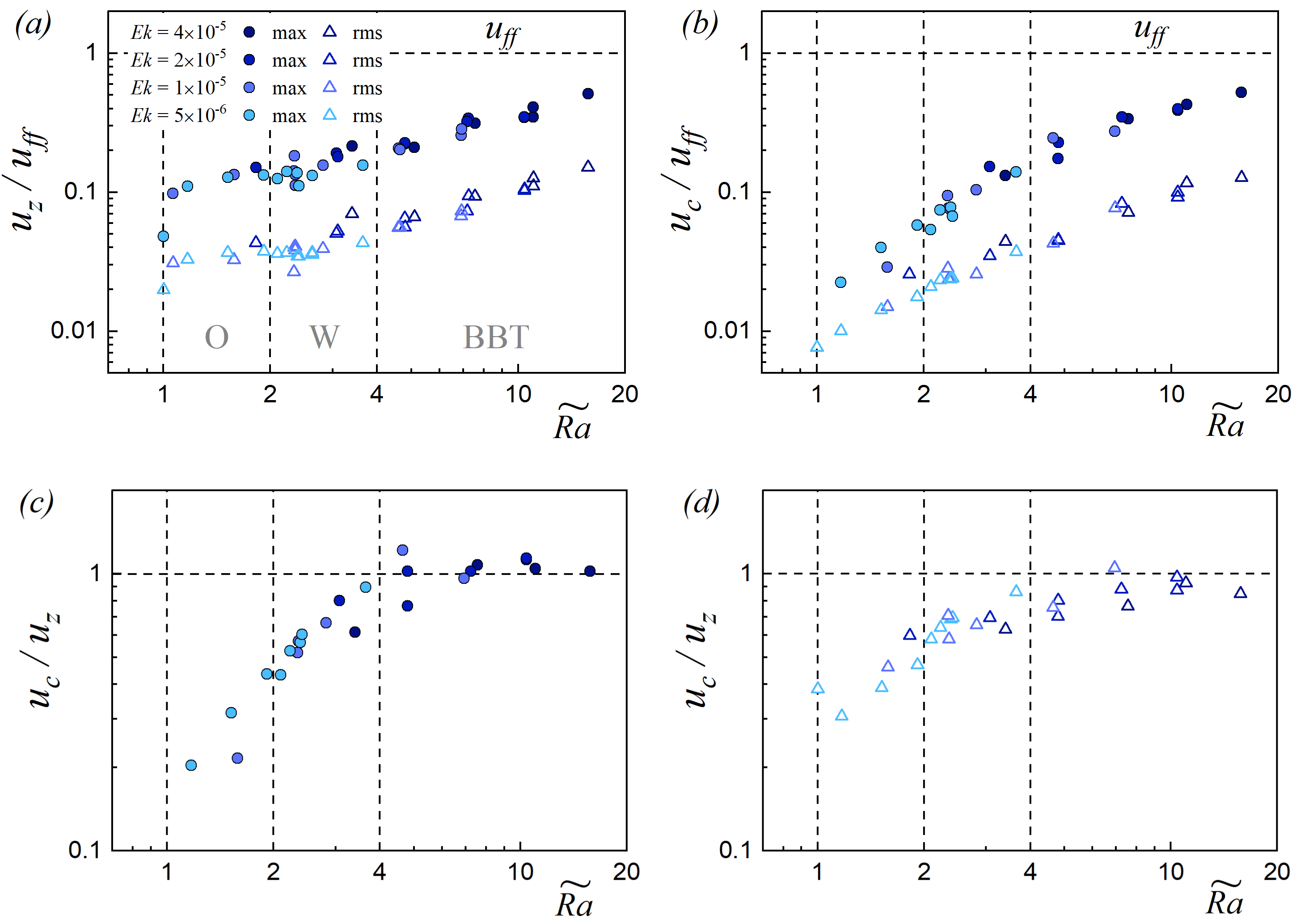}
 \caption{UDV measurements of maximum (circle) and root mean square (triangle) velocities as a function  of convective supercriticality $\widetilde{Ra} = Ra/Ra_O^{cyl}$. (a) Vertical velocities, $u_z$, depth-averaged over $0.45 \leq z/H \leq 0.55$. (b) Chord velocities, $u_c$, spatially-averaged over $0.45 \leq c/C \leq 0.55$. \TV{(c) and (d) shows the ratio of the chord velocity and vertical velocity for (c) maximum values and (d) root mean square values. (c) and (d) shows that the vertical and horizontal kinetic energies are roughly similar in the broad band turbulence regime.}}
\label{fig6}
 \end{figure}

Figure \ref{fig6} compares vertical and chord probe velocities, $u_z$ and $u_c$, respectively. 
The filled circles show maximum velocity values and the triangular symbols mark the root mean square values. The velocity values were each averaged in an approximately 10 mm wide measuring window centered at the half height of the cylinder ($0.45 \leq z/H \leq 0.55$) or around the midpoint of the chord ($0.45 \leq c/C \leq 0.55$).   If the vertical oscillations are described by a sinusoidal oscillation (as is expected at onset \citep{Chandrasekhar1961, Julien1998}), then the maximum and rms values would be related via $u_{z,max} = \sqrt{2} \, u_{z,rms}$. This, however, is not the case. The typical factor between the maximum and rms values is closer to 3.5 (see table \ref{tab:kd}), implying that these are no longer simple sinusoidal oscillations. The same holds for the chord velocity values. 

\JA{Figures \ref{fig6}(c,d) show that the vertical and horizontal velocity magnitudes are of the same order of magnitude across our entire range of experiments, and are within $\simeq 30\%$ of one another in the broadband turbulence regime.}  This equipartitioning of vertical and horizontal kinetic energies, $u_z^2 \sim u_c^2$, in our low $Pr$, $Ra < Ra_S^\infty$ oscillatory cases is qualitatively similar to the equipartitioning found in the $Pr \sim 1$ non-magnetic rotating convection DNS of \citet{Stellmach2004} and \cite{Horn2015} and in the rapidly rotating asymptotically-reduced models of \citet{Julien2012gafd}.

 \subsection{Spectra} 
\begin{figure}
\includegraphics[width=\textwidth]{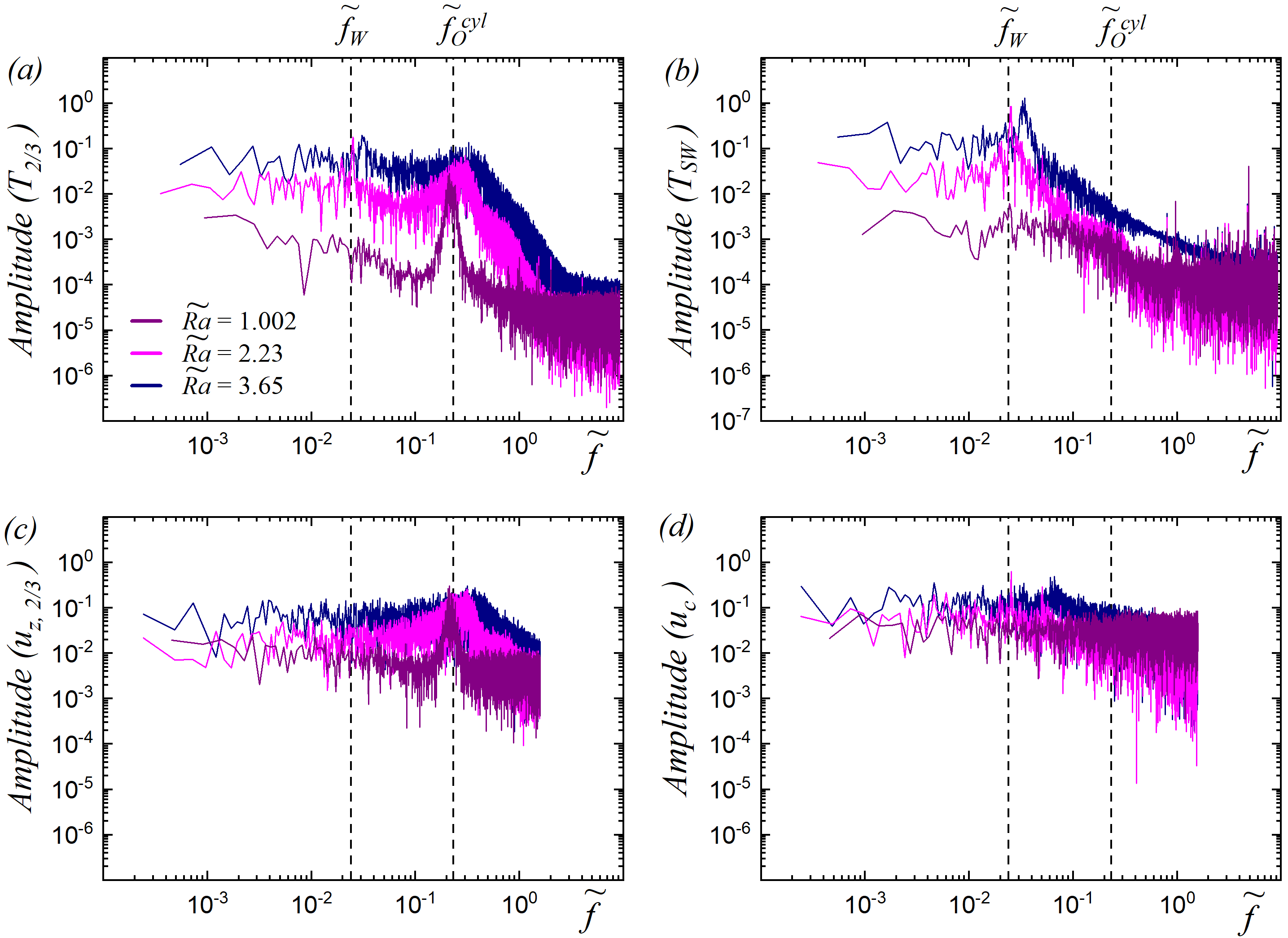}
\caption{\TV{Amplitude of the }Fourier transforms of temperature and velocity signals versus normalized frequency $\widetilde{f} = f / f_{\Omega}$. The Ekman number is $Ek=5 \times 10^{-6}$ and the supercriticality $\widetilde{Ra}$ is indicated by the line color. All spectra are evaluated on the midplane, $z/H=1/2$. (a) Temperature spectra measured with a thermistor situated within the fluid bulk at $r/R = 2/3$. (b) Temperature spectra measured on the cylindrical tank's outer sidewall at $r/R = 1.05$. (c) Vertical velocity spectra measured at $r/R = 2/3$. (d) Chord velocity spectra evaluated in the vicinity the the sidewall. Vertical dashed lines indicate the onset frequency for wall modes $\widetilde{f}_W=0.024$ 
and bulk oscillations $\widetilde{f}_O^{cyl}=0.274$.}
\label{fig7}
 \end{figure}

Spectral analyses of the temperature and velocity time series allows us to characterize the spatiotemporal modal content of our experimental results. First, we will look at the temperature spectra. The temperature spectral analysis concentrates on two thermistors located at half height of the fluid layer, $z = H/2$. One thermistor, denoted $T_{2/3}$, is located in the fluid bulk at a radial position $r/R \approx 2/3$. The other is attached to the exterior of the stainless steel sidewall of the vessel at $r/R=1.05$, and is denoted $T_{SW}$. The spectra of these thermistors are shown in figure \ref{fig7}(a-b) for a constant $Ek = 5 \times 10^{-6}$ and $\widetilde{Ra} = (1.002, 2.23$ and 3.65). The spectrum at the lowest supercriticality case at $\widetilde{Ra} = 1.002$ shows a clear peak at the predicted value $\widetilde{f}^{cyl}_{O}$ on thermistor $T_{2/3}$ (figure \ref{fig7}a). In contrast, the spectrum at the sidewall, shown in figure \ref{fig7}(b), does not show a pronounced peak for $\widetilde{Ra} = 1.002$. The spectrum at $\widetilde{Ra} = 2.23$ reveals a clear peak on both themistors at the predicted wall mode frequency $\widetilde{f}_W$. The peak related to the oscillatory convection has become broader and shifted towards higher frequencies. The spectrum at $\widetilde{Ra} = 3.65$ shows further evidence for wall modes and oscillatory convection in figure \ref{fig7}(a). Both peaks have shifted towards higher frequencies with respect to the predicted frequency at onset. Additionally, the width of the frequency peak around $\widetilde{f}^{cyl}_{O}$ has further expanded. A comparison of figure \ref{fig7}(a) and (b) shows that the fluid bulk is dominated by oscillatory convection whereas the wall modes remain dominant in the vicinity of the sidewall.

Figures \ref{fig7}(c,d) show the corresponding velocity spectra, which are evaluated at two locations comparable to the thermistor positions in figures \ref{fig7}(a,b). The velocity spectra in figure \ref{fig7}(c) were recorded with the UDV transducer that measures the vertical velocity at a radial position $r/R = 2/3$. The velocity data are depth-averaged over $0.45 < z/H < 0.55$. A comparison of the corresponding figures \ref{fig7}(a) and (c) reveals a qualitative agreement of the peak frequency $\widetilde{f}^{cyl}_{O}$. In contrast, wall mode peaks are evident in the $T_{2/3}$ spectra but not in the $u_{z,2/3}$ spectra. These wall mode signatures are visible in the temperature spectra and not in the velocity spectra due to the low $Pr$ nature of the fluid. \SH{All the wall modes signatures exponentially decay inwards from the sidewall, but the thermal signatures extends much farther into the fluid bulk \JA{than the velocity signatures since $\kappa \simeq 40 \, \nu$ in gallium} \citep[][]{Horn2017, Aurnou2018}.}

%

Figure \ref{fig7}(d) shows velocity spectra recorded with the chord probe and evaluated in the vicinity of the sidewall ($0.05 \leq c/C \leq 0.07$). The $u_c$ spectra show evidence for wall modes, although they are much less well pronounced than in the temperature spectra. 

The relative height that a fluid element travels during one convective oscillation can be roughly approximated by assuming a sinusoidal vertical motion, $\delta z/H = u_{z,rms}/(2 f_O^{cyl} H)$. We take characteristic values of $f_O^{cyl} = 0.125$ Hz and $u_{z,rms} \simeq 2.2$ mm/s for the experiments in the oscillatory regime $1 \leq \widetilde{Ra} \leq 2$.  This gives a relative travel distance of $\delta z/H  \approx 0.09$, such that the fluid traverses roughly 10\% of the fluid layer depth over its oscillatory path. Since this implies that thermal anomalies are not advected vertically across the entire layer, these oscillatory modes should be relatively inefficient in transporting heat (see figure \ref{fig4}(a)). Figure \ref{fig4}(b) and figure \ref{f:scalings}(a) shows that, even though the low $Pr$ oscillatory modes are thermally inefficient, they generate relatively high flow velocities that approach $u_{f\!f}$ even at relatively low supercriticalities (e.g., $Ra < Ra_S^\infty$).
\begin{figure}
\includegraphics[width=\textwidth]{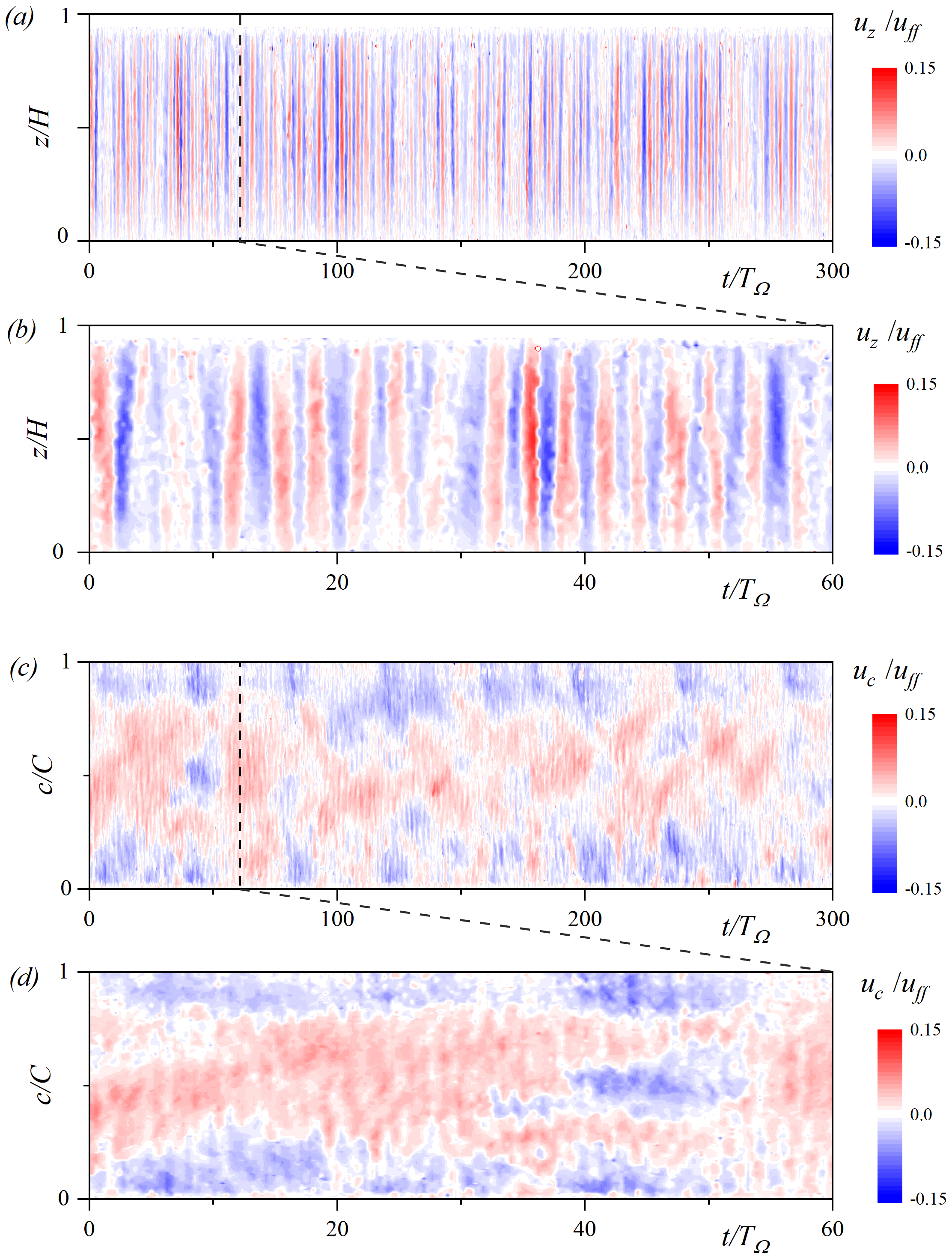} 
 \caption{UDV Dopplergrams: Spatiotemporal evolution of laboratory convection velocities at $Ek =  5 \times 10^{-6}$ and $\widetilde{Ra}=2.23$ ($Ra=7.7 \times 10^6$, $Pr=0.026$, $\Gamma=2$). (a, b) vertical velocity distribution along the cylinder height, where positive (red) values correspond to upwards directed flows. (c,d) Velocity distribution along the chord, where positive values correspond to flows in the direction of rotation (prograde). All velocity values are normalized by the free-fall velocity $u_{f\!f} = \sqrt{\alpha g \Delta T H} = 59$ mm/s for this case.}
\label{fig8}
\end{figure} 
 \begin{figure}
\includegraphics[width=\textwidth]{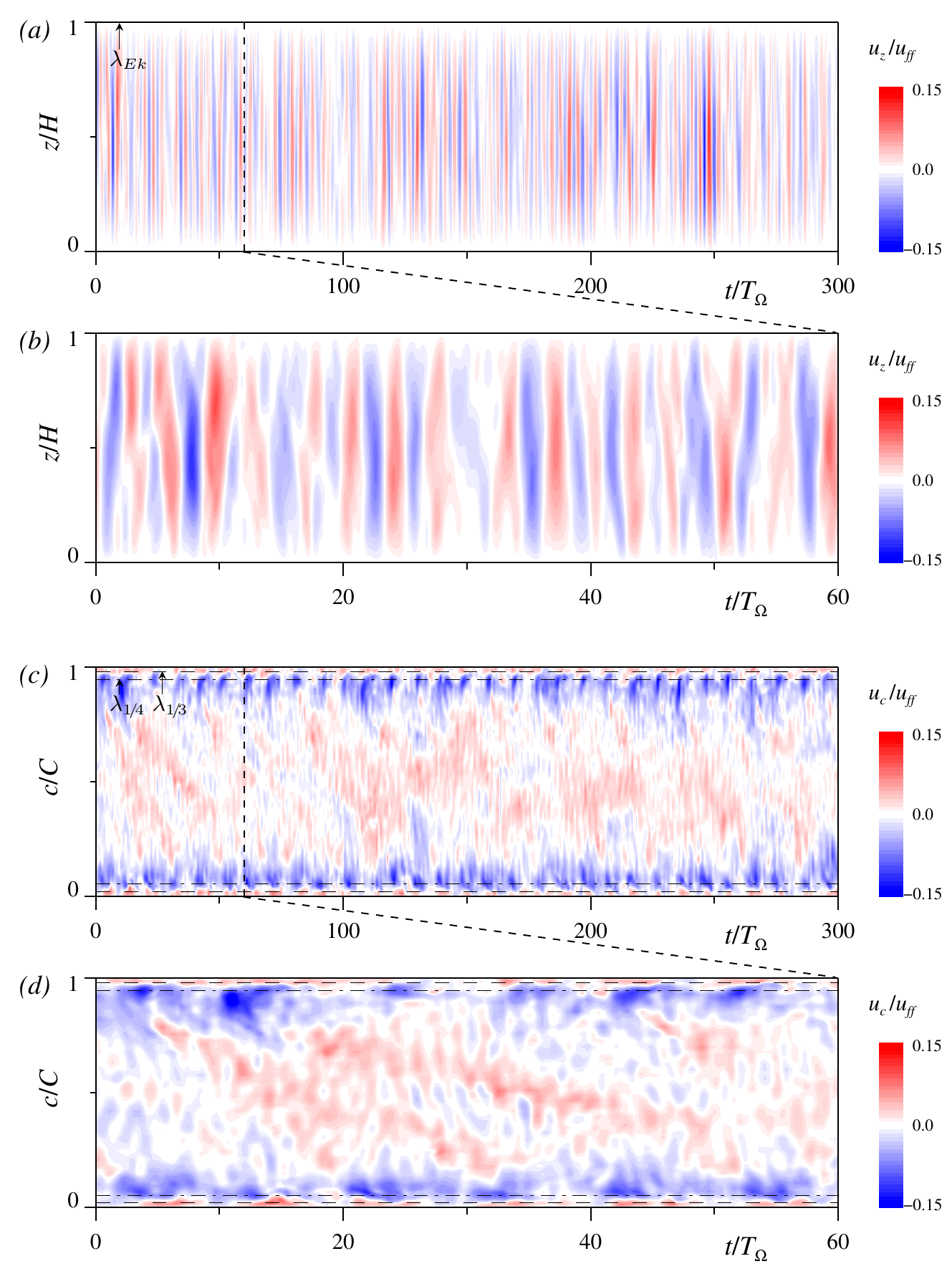}
\caption{Synthetic Dopplergrams obtained from the $\widetilde{Ra} = 2.30$ DNS ($Ek = 5 \times 10^{-6}$; $Ra = 8.0 \times  10^6$; $Pr=0.025$; $\Gamma=1.87$). The visualization scheme is identical to that of figure \ref{fig8}. The thickness of the Ekman boundary layer is marked by the dashed $\lambda_{Ek}$ line just below the upper boundary in (a). The thicknesses of the inner and outer Stewartson boundary layers are indicated by the dashed $\lambda_{1/3}=(2 \, Ek)^{1/3}$ lines and dot-dashed $\lambda_{1/4}=(2 \, Ek)^{1/4} $ lines in (c) and (d). All velocity values are normalized by the free-fall velocity $u_{f\!f} = \sqrt{\alpha g \Delta T H} = 61$ mm/s for this case.}
\label{fig9}
\end{figure}

\subsection{Canonical case \label{subsec:canonical}}
In this section, we focus on the flow field in the $\widetilde{Ra} = 2.23$ laboratory case and the corresponding $\widetilde{Ra} = 2.30$ DNS, both of which are in the multimodal regime with coexisting bulk oscillations and wall-modes. The laboratory Dopplergrams presented in figure \ref{fig8}(a,b) show the evolution of the vertical velocity for different time frames. The ordinate is normalized by the fluid layer depth $H$, the abscissa is normalized with the system's rotation time $T_\Omega$ and the velocity color scale is normalized by the free fall velocity $u_{f\!f}$. A positive value of $u_z/u_{f\!f}$ represents an upwardly directed flow.  A regular oscillation over the entire cylinder height is clearly visible. The amplitude of the oscillation is already in the range of $u_z/u_{f\!f} \approx \pm 0.15$, although this case is still moderately close to onset. The oscillating axial velocity field differs significantly from the stationary columns that appear in rotating convection at $Pr \geq 1$ fluids, where \TV{coherent unidirectional flows} in the $z$-direction are observable \citep{Stellmach2014}. 

Figure \ref{fig8}(c,d) show Dopplergrams of the chord velocity data. The measuring depth, displayed on the ordinate, is normalized with the total length of the chord $C$. The chord probe measures a combination of the cylindrically radial velocity $u_r$ and the azimuthal velocity $u_{\phi}$ (section \ref{tech}). The time-averaged velocity distribution along the chord is approximately symmetrical around $c/C=0.5$, which leads to the conclusion that the mean radial flow component is negligible. This quasi-symmetry of the $u_c$ profiles was observed in all measurements above $\widetilde{Ra} \geq 2$. At the chord's midpoint $c/C=0.5$ ($r = 0.7R$), the UDV senses only the azimuthal velocity component, $|u_c| = |u_{\phi}|$. At the other measuring depths $|u_c| < |u_{\phi}|$ is valid because only the projection of $u_{\phi}$ on the chord is captured by the sensor.

The flow field near the sidewall, $c/C \approx 0;1$, is dominated by retrograde (opposite direction as the applied $\Omega$)  azimuthal flows ($u_c < 0$). The middle range of the chord is dominated by prograde (same direction as the applied $\Omega$) velocities ($u_c > 0$). The range of velocities in the chord direction is comparable to the velocity range in vertical direction. When looking at the near wall region of the chord measurement, a low frequency oscillating behaviour becomes visible and indicates the presence of wall-modes (cf.~figure \ref{fig7}b). The frequency of the wall modes is $\widetilde{f}_W=0.025$, which compares well with \citet{Zhang2009}'s predicted value of $\widetilde{f}_W=0.024$. 

Figure \ref{fig9} shows synthetic numerical Dopplergrams, constructed similarly to figure  \ref{fig8}. Figures \ref{fig8} and \ref{fig9} show good agreement between laboratory and DNS results, with both containing oscillatory, coherent axial up- and downwelling velocities in the bulk as well as wall modes on the lateral periphery of the domains. 
Noteable differences exist, however, in the regions near the side wall. Both physical and metrological causes are considered for these differences. The boundary conditions between experiment and the DNS have minor differences in the aspect ratio and in the thermal boundary conditions. While the side walls are thermally perfectly insulated in the numerical simulation, low radial and axial heat fluxes through the stainless steel wall occur in the experiment. Deviations in the thermal boundary conditions may influence the wall mode properties, since wall modes are sensitive to the thermal boundary conditions \citep[e.g.,][]{Herrmann1993}. \TV{A possible metrological \JA{explanation}} for the qualitative deviations between figures \ref{fig8}(c,d) and figures \ref{fig9}(c,d) is the \JA{differing measuring volumes used in the UDV and the DNS measurements.} The UDV measurements have a cylindrical measuring volume which has a beam diameter of about 5 mm. The finite diameter of the UDV measuring volume leads to a radial averaging of the velocities, especially in areas close to the curved side wall. In contrast, the synthetic numerical Dopplergrams measure the velocity distribution along an ideal line \JA{of zero width}. As a result, the wall modes are more pronounced in the numerical Dopplergrams in comparison to the UDV measurements. 

The dashed horizontal line right at the top of \ref{fig9}(a) shows the thickness of the Ekman boundary layer, $\lambda_{Ek}   = Ek^{1/2} H \simeq 2 \times 10^{-3} H$.
The dashed and dot-dashed horizontal lines in \ref{fig9}(c,d) show the respective inner and outer Stewartson boundary layer thicknesses on the cylindrical sidewall 
\begin{equation}
\lambda_{1/3}=(2 \, Ek)^{1/3} \,H \,\,  \mbox{ and } \,\, \lambda_{1/4}=(2 \, Ek)^{1/4} \, H 
\label{eq:stew1}
\end{equation}
\citep{Stewartson1957, Friedlander1980, Kunnen2013}. These Stewartson layer predictions are in good qualitative agreement with the locations of the sidewall flow structures. 

 \begin{figure}
\includegraphics[width=\textwidth]{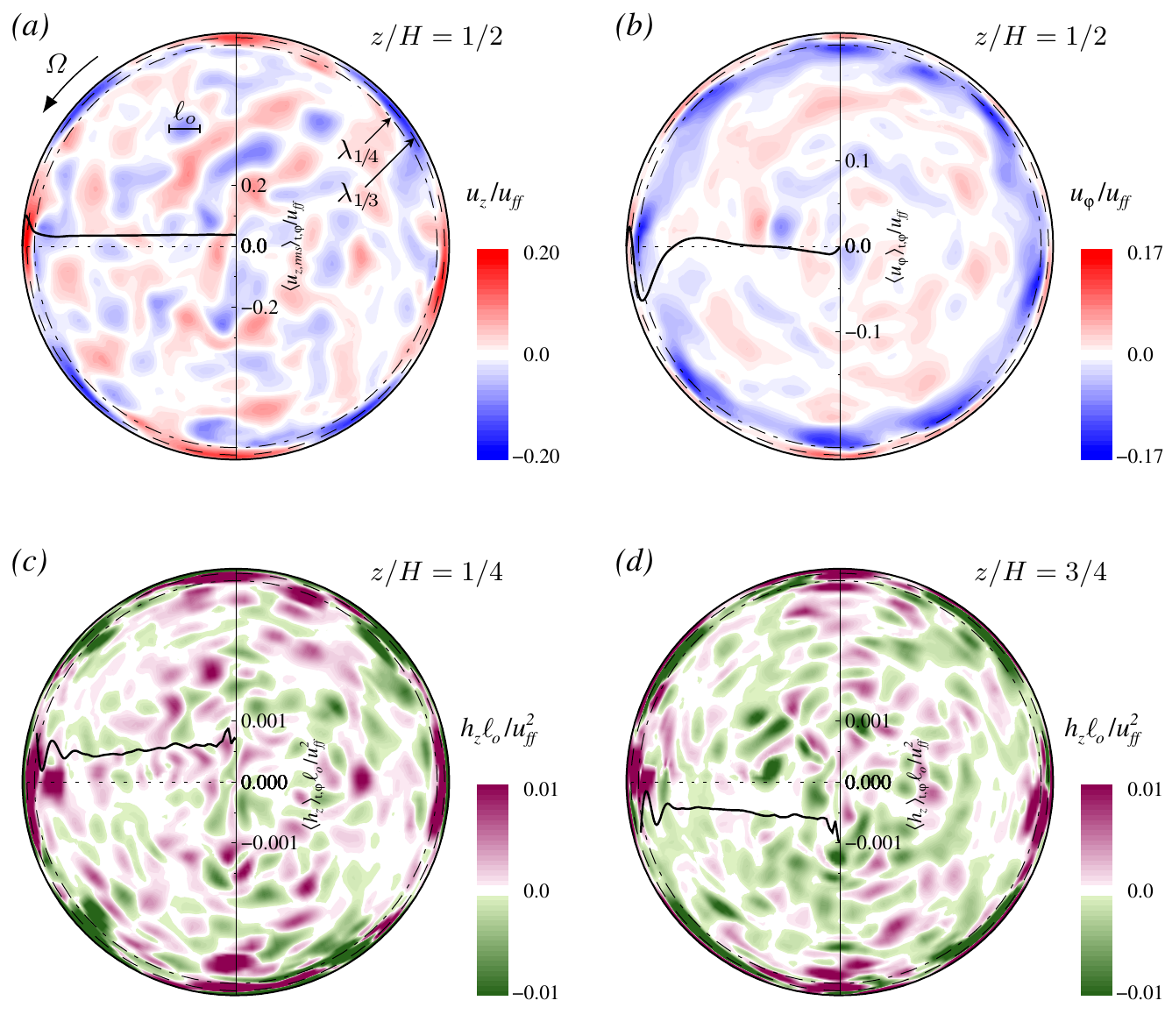}
 \caption{DNS: \SH{Normalised instantaneous velocity and helicity distributions on different horizontal cross-sections. The inner and outer Stewartson layers, $\lambda_{1/3}$ and $\lambda_{1/4}$, are marked with a dashed and dot-dashed line, respectively; $\ell_O^\infty$ indicates the oscillatory onset length scale according to eq.~\eqref{eq:length}.} (a) vertical velocity $u_z$ at half height; (b) azimuthal velocity $u_{\phi}$ at half height;
(c,d) helicity $h_z = u_z \omega_z$ at $z/H = 1/4$ and 3/4, respectively.  The black curves in each panel show the time-azimuthal mean of that quantity. In the near wall region, the normalized helicity swings between -0.032 and 0.014 on the $z/H = 1/4$ plane. Comparable helicity values, with signs \SH{reversed}, are found on the $z/H = 3/4$ plane.} 
\label{fig10}
 \end{figure}

Figure \ref{fig10} shows velocity and helicity data along different horizontal cross-sections of the $\widetilde{Ra} = 2.30$ DNS. Figure \ref{fig10}(a) shows a snapshot of the vertical velocity distribution in a horizontal section at half height of the cylinder, $u_z(r, \phi, z = H/2) / u_{f\!f}$. A positive (red) velocity corresponds to an upward flow. The solid black line in the left half of the image shows the radial profile of the time and azimuthal root mean square (rms) velocity, showing an almost constant value in the bulk with a maximum located near the side wall. The theoretical predicted length scale of the vertical oscillations, $\ell_O^\infty$, is shown in the upper left quadrant. Good agreement exists between this theoretical length scale to the observed size of the flow structures in the bulk. The boundary between the oscillation-dominated-bulk and the wall-mode-dominated sidewall region correlates well with the $\lambda_{1/4}$ outer Stewartson boundary layer (dot-dashed circle). The maximum vertical velocity, on the other hand, correlates well with the $\lambda_{1/3}$ inner Stewartson layer (dashed circle). The wall modes have an azimuthal wave number of $m = 4$ based on this $u_z$ data. 

Figure \ref{fig10}(b) shows a snapshot of the midplane azimuthal flow field $u_{\phi}(r, \phi, z = H/2) / u_{f\!f}$. The solid black line in the left half of the image shows the time and azimuthal averaged velocity profile. The azimuthal flow field is mainly dominated by wall modes. The outermost regions show a prograde flow whose velocity maxima correlate approximately with the $\lambda_{1/3}$ inner Stewartson layer. Further away from the sidewall, there exists a retrograde flow whose velocity minima correlate approximately with the $\lambda_{1/4}$ outer Stewartson layer. The radial width of the wall modes in the azimuthal flow field is larger than in the vertical velocity field, in good agreement with \citet{Kunnen2013}. The wall mode wave number appears to be $m = 8$ in the azimuthal velocity data, but this is a fallacy.  The $u_\phi$ component of the wall mode is antisymmetric with respect to the midplane, resulting in an effective phase shift of half a wavelength between the upper and lower hemicylinder. Due to the nonlinearity of the wall modes, however, signatures of the wall mode vortices from both hemicylinders are observable on the midplane. This results in this apparent, but not factual, wave number doubling in figure \ref{fig10}(b). That nonlinear wallmodes do not have a sharp antisymmetry across the midplane was observed previously in \citet{Horn2017} (see their figure 9(b) and supplementary movies 7 and 11).

Figures \ref{fig10}(c) and (d) show snapshots of the non-dimensional vertical helicity, 
$h_z = u_z \omega_z$, in the respective midplanes of the lower hemicylinder ($z/H=1/4$) and of the upper hemi-cylinder ($z/H=3/4$).  The time-azimuthal mean radial profiles of bulk vertical helicity, $\langle h_z \rangle_{t,\phi}$, are represented by the solid black lines in the left half of each image. 
The profiles extend from $r=0$ to $r = R - \lambda_{1/4}$, since the side wall flows generate far stronger local $|h_z|$ signals that swamp the values in the fluid bulk. The bulk $\langle h_z \rangle_{t,\phi}$ profiles show that low $Pr$ oscillatory convection generates mean negative helicity in the lower hemi-cylinder and mean positive helicity in the upper hemicylinder.  This is qualitatively visualized in panels (c) and (d), where positive (purple) patches dominate in the lower hemicylinder shown in (c) and negative (green) patches dominate in the upper hemicylinder shown in (d). 

\JA{This hemicylindrical time-mean helicity distribution is qualitatively similar to the rotating magnetoconvection modeling results of \citet{Giesecke2005} and  the rotating convection models of \citet{Schmitz2010}.  More quantitatively, we find, using the water simulation with $Pr = 6.4$, $Ek = 2 \times 10^{-4}$, $Ra = 2.6 \times 10^6$, $\Gamma = 2$ from \citet{Horn2017}, that the time-mean helicity is roughly 10 times greater than that of the turbulent liquid metal DNS presented here. The $Pr = 6.4$, quasi-laminar ($Re_{z, max} = 115$; $Re_{\ell_S^\infty} = 15$; $Ro_{\ell_S^\infty} = 0.18$; $Ra = 2.7 Ra_S^\infty$) helicity values are likely comparable to the more turbulent ($Re_{z, max} = 2414$; $Re_{\ell_O^\infty} = 325$;  $Ro_{\ell_O^\infty} = 0.09$; $Ra = 2.23 Ra_O^{cyl}$)  liquid metal DNS when one considers that the steady columns tend to lose their axial coherence in the  quasi-geostrophic turbulence regime \citep{Julien2012gafd, Aurnou2015}.  It remains an open question as to how the helicity varies as a function of $Pr$ in rapidly rotating strongly turbulent convection.}

\begin{figure}
 \includegraphics[width=\textwidth]{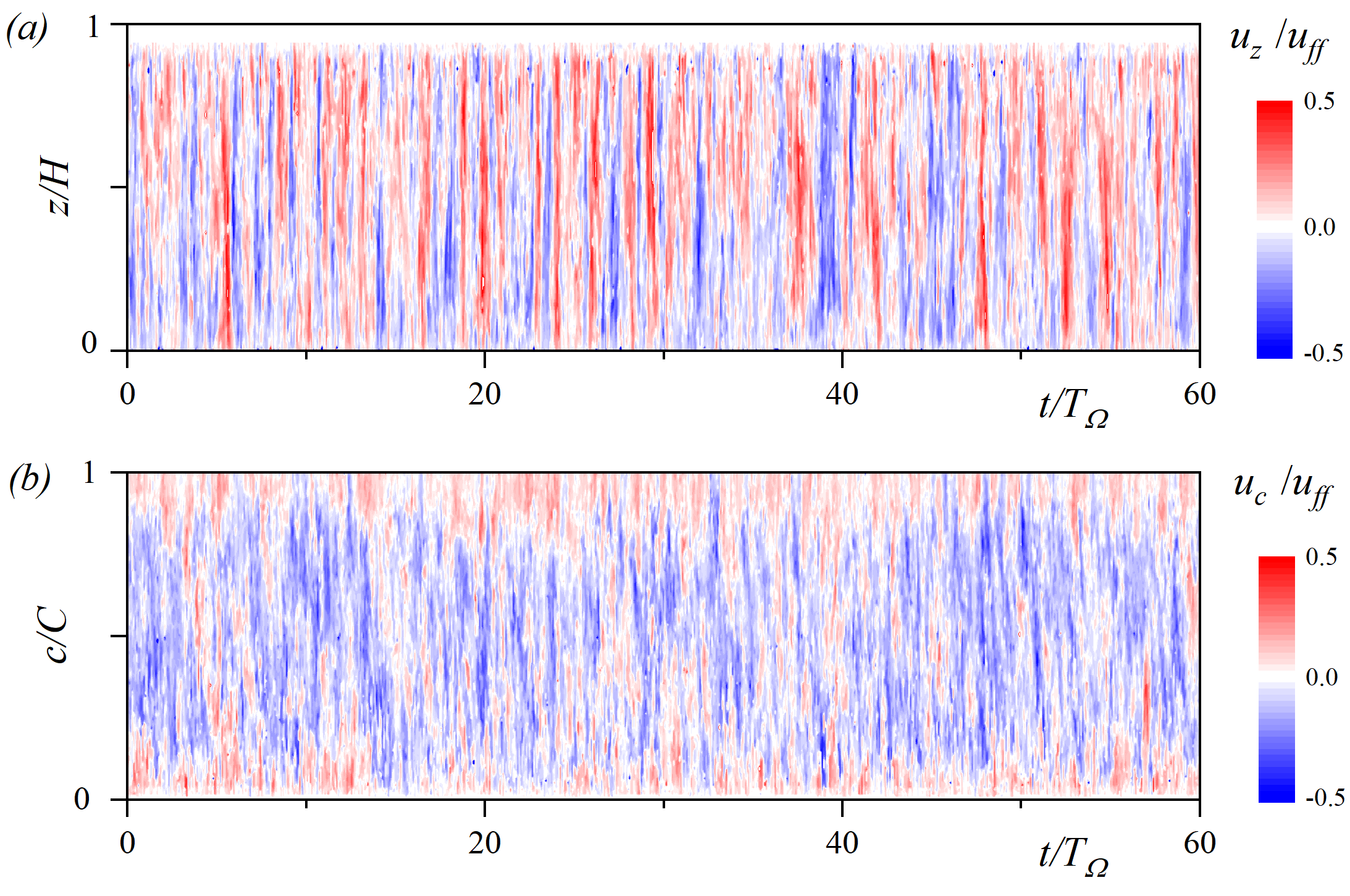}
 \caption{UDV Dopplergrams for laboratory high supercriticality case at $\widetilde{Ra}=15.7$ ($Ek =  4 \times 10^{-5}$, $Ra=4.4 \times 10^6$, $Pr=0.026$, $\Gamma=2$). (a) vertical velocity distribution along the cylinder height; (b) velocity distribution along the midplane chord. All velocity values are normalized by the free-fall velocity $u_{f\!f} = \sqrt{\alpha g \Delta T H}=45.1$ mm/s in this case.}
\label{fig11}
 \end{figure}

The DNS flow field analyses support the idea of a fully self-consistent dynamo in the oscillatory regime, as has also been proposed by \citet{Calkins2015, Calkins2016b, Davidson2018}.  Helicity is an important ingredient in many dynamo systems \citep[e.g.,][]{Schmitz2010, Soderlund2012}, where it is well known that north-south hemispherical dichotomies in the helicity field can produce axial dipole dynamo solutions \citep[e.g.,][]{McFadden1988, Davidson2015, Moffatt2019}. The hemicylindrical helicity dichotomy found here is not a time-varying phenomenon. \JA{Therefore, low $Pr$ oscillatory rotating convective flows contain the essential kinematic ingredients for system-scale magnetic field generation.}


\subsection{Strongly supercritical case}

Figure \ref{fig11} shows UDV Dopplergrams for our highest supercriticality case ($\widetilde{Ra} = 15.7$).  The vertical velocity field in figure \ref{fig11}(a) is still dominated by vertical oscillations that nearly extend over the entire fluid layer depth, even though the parameters for this case are relatively far above onset. The frequency of the oscillations is, however, significantly higher than in cases closer to $Ra_O^{cyl}$ (cf. figure \ref{fig8}b), and their appearance is less regular in both space and time \TV{\citep{Julien1998, Horn2017, Aurnou2018}}. 
The flow velocities reach $\approx 50\%$ of the free fall velocity in both measured directions, and $\approx 95\%$ of the velocity scaling predictions from the liquid gallium RBC experiments of \citet{Vogt2018jrv}. The oscillatory convection velocities in our experiments correspond to maximum Reynolds number of order $Re \approx 7 \times 10^3$, and are likely fully turbulent even though $Ra \lesssim 0.7 Ra_S^\infty$.
The chord probe data, shown in figure \ref{fig11}(b), shows a strong prograde motion close to the sidewall, whereas, on average, a retrograde flow exists in the bulk. Thus, the azimuthal flow field has become inverted relative to that in figure \ref{fig9}.  The chord probe velocity range and time scales are comparable to those in the vertical velocity data, similar to the lower $\widetilde{Ra}$ data in figures \ref{fig8} and \ref{fig9}.
 
\subsection{Zonal flows}
\begin{figure}
 \includegraphics[width=\textwidth]{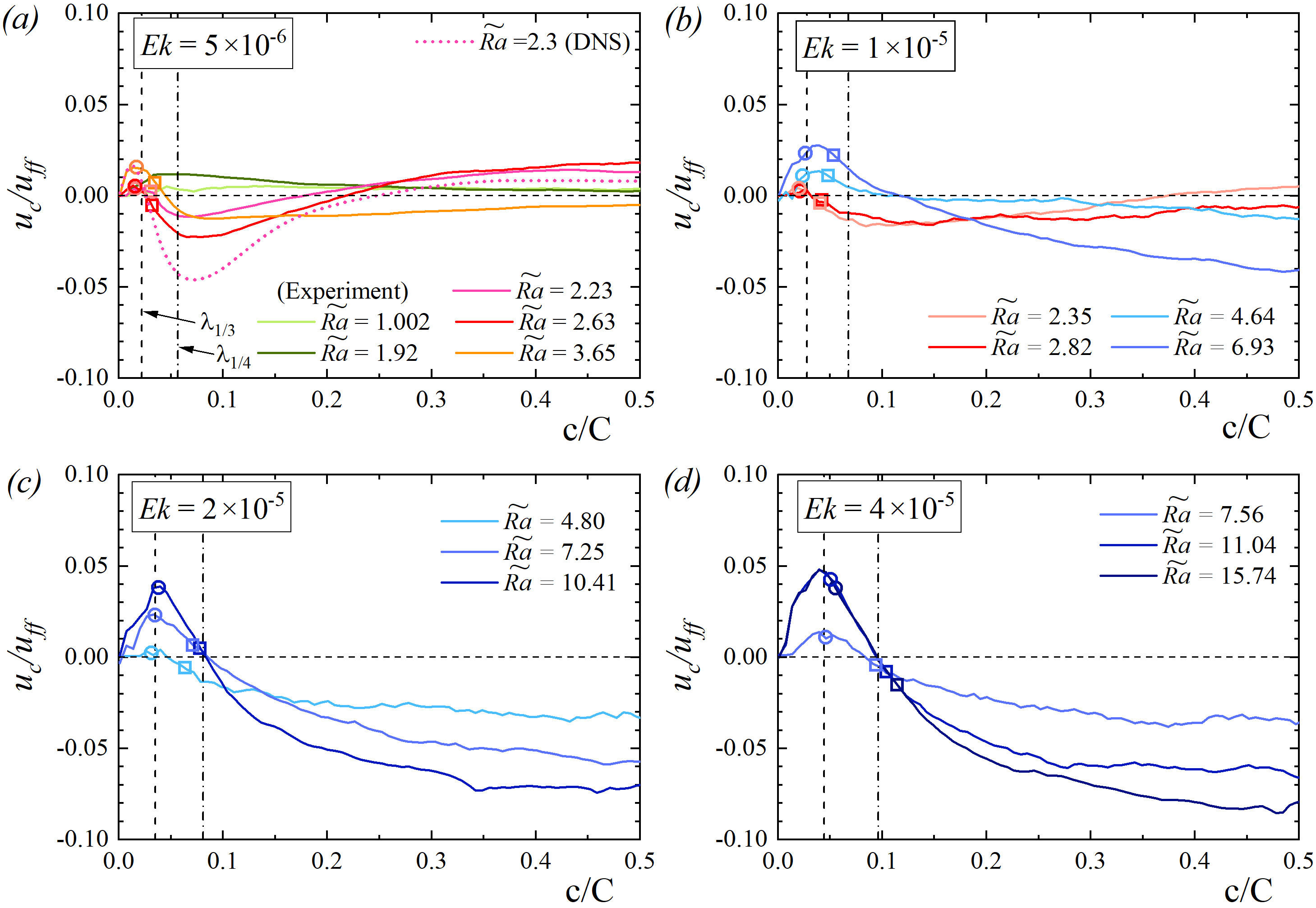}
 \caption{Time averaged chord probe velocity profiles, $u_c$, for different $\widetilde{Ra}$ and $Ek$. The solid line represents laboratory UDV  data and the dotted line in panel (a) corresponds to the $\widetilde{Ra}=2.30$ DNS. Positive values correspond to flow into the direction of rotation (prograde); the vertical dashed lines indicate the thickness of the inner and outer Stewartson layers, $\lambda_{1/3}$ and $\lambda_{1/4}$, projected onto $c/C$ coordinates. The open circles indicate the proposed boundary zonal flow (BZF) scaling $Ra^{1/4}Ek^{2/3}$ and the squares indicate a slightly modified $Ra^{1/4}(2Ek)^{2/3}$ scaling.}
\label{fig12}
 \end{figure}

Figure \ref{fig12} shows time-averaged chord probe UDV profiles, $u_c$, for different $\widetilde{Ra}$ and $Ek$. The velocity distribution is plotted over the measuring distance $0 < c/C < 1/2$. The accuracy of the velocity profiles at the end of the measuring line at $c/C=1$ is affected by multiple reflections of the ultrasound signal on the curved side wall. For this reason, we plot only the first half of the approximately symmetrical profiles in figure \ref{fig12}. Panel (a) through (d) shows cases at successively higher $Ek$, corresponding to higher $\widetilde{Ra}$ conditions.  The color scale of each profile is selected such that green hues denote the oscillatory ($O$) regime; red, pink and orange hues denote the wall mode dominated ($W$) regime; and blue hues denote cases in the broad band ($BBT$) regime. The vertically dashed and dashed-dotted lines indicate the inner and outer Stewartson  layers, respectively. 

Figure \ref{fig12}(a) displays the $u_c$ velocity profiles in the $Ek = 5 \times 10^{-6}$ cases, which correspond to our lowest $\widetilde{Ra}$ experiments. The green-hued curves, corresponding to the oscillatory regime, show a weak prograde flow. The red-hued curves, corresponding to the wall mode regime, show increasing $|u_c|$ values. The $\widetilde{Ra}=2.30$ DNS case's synthetic $u_c$ profile is demarcated via the dotted pink curve.   The velocity amplitudes in the DNS are larger compared to the experimentally determined profiles, especially near the sidewall ($c/C \lesssim 0.15$).  This difference is likely due to the thermal sidewall boundary conditions and the finite width of the ultrasonic beam, which leads to an averaging effect that is especially strong near the cylindrical sidewall (cf.~\textsection\ref{subsec:canonical}). 

The maximum of the prograde flow occurs near the wall, and is located in the vicinity of the $\lambda_{1/3}$ inner Stewartson layer, in good agreement with \citet{deWit2020, Favier2020, Zhang2020}. In lower $\widetilde{Ra}$ cases, a velocity minimum is located near to the predicted $\lambda_{1/4}$ outer Stewartson layer. In the bulk there is a prograde flow observed in lower $\widetilde{Ra}$ cases, but it changes into a retrograde flow in the $\widetilde{Ra} \geq 3.65$ cases.  Qualitatively, the velocity distributions in the broad band regime ($\widetilde{Ra} \gtrsim 4$) are all similar, showing a maximum located at $\lambda_{1/3}$ and a strong retrograde bulk flow whose intensity increases with $\widetilde{Ra}$. The zero crossing point in the velocity profiles between the prograde flow at the side wall and the retrograde flow in the interior may correlate with $\lambda_{1/4}$ in the higher supercriticality cases shown in figures \ref{fig12}(c) and \ref{fig12}(d).

The strong side wall circulations in figure \ref{fig12} are called boundary zonal flows (BZFs), following \citet{Zhang2020}\SH{. They have been shown to be linked to non-linear wall modes processes and Stewartson boundary layer like behaviours \citep{deWit2020, Favier2020, Shishkina2020}.}
Our experiments are the first to show the existence of BZF in low $Pr$ fluid, and in the regime where steady convective modes are not yet viable ($Ra < Ra_S^\infty$). Thus, our results support the notion that BZFs are a universal feature of rotating convective flows. \citet{Zhang2020} argued that the zero-crossing, i.e. the first instance away from the sidewall where $\langle u_\phi \rangle_{t,\phi} =0$, is a measure of the radial thickness of the BZF. In their $Pr \simeq 0.8$. experiments in SF\textsubscript{6}, they found that the BZF defined as such scales as $Ra^{1/4}Ek^{2/3}$; the entire BZF becomes thicker with increasing $Ra$ at constant $Ek$. This thickening of the BZF has been associated with wall modes becoming non-linear \citep{Favier2020}.  We demarcated \citet{Zhang2020}'s proposed $Ra^{1/4}Ek^{2/3}$ BZF scaling by the circles in figure \ref{fig12}. In our experiments this scaling always lines up well with the peak prograde location of the BZF, near $\lambda_{1/3}$.  We argue that this alignment likely occurs since $\widetilde{Ra} = \mathcal{O}(1)$ in our experiments. \JA{Taking $Ra \sim Ra_c$ yields}
\begin{equation}
Ra^{1/4}Ek^{2/3} \sim \left((Ek/Pr)^{-4/3} \right)^{1/4}Ek^{2/3} = (Ek\, Pr)^{1/3} \sim \lambda_{1/3}, 
\label{BZFscaling}
\end{equation}
which implies that the $Ra^{1/4}Ek^{2/3}$ scales with the inner Stewartson layer thickness near onset.  

As an alternative, the square symbols in figure \ref{fig12} denote $Ra^{1/4}(2Ek)^{2/3}$, based on the argument that sidewall boundary thicknesses are usually better captured via a function of $\nu/(\Omega H) = 2 Ek$ \citep{Stewartson1957,Kunnen2013}. This second scaling only captures the zero crossings for higher values of $Ek$; in the rapidly rotating, low $Ek$ limit, these two scalings will tend to converge towards $\lambda_{1/3}$ as shown in (\ref{BZFscaling}). Thus, it is not clear from our results that \JA{the existing scaling predictions adequately describe} the scaling properties of the BZF zero-crossing points in our experimental data.  Note that the $Pr^{1/3}$ term in (\ref{BZFscaling}) exists only in low $Pr$ fluids. \JA{This suggests that experiments in which $Pr$ is strongly varied will be capable of determining the scaling behaviours of the BZF.}

The strong retrograde azimuthal bulk flow in the $\widetilde{Ra} > 4$ cases may also indicate the formation of a large scale vortex (LSV). LSVs can develop by an inverse cascade in which energy is transported from small to large scales. The occurrence of an inverse cascade is favored by a two-dimensionality of the large-scale flow field \citep{Kraichnan1967, Boffetta2012}, which can arise in geo- and astrophysical systems through the action of stabilizing forces such as the Coriolis force due to rotation or via Lorentz forces induced by magnetic fields. In rotating convection, LSV structures can form if there is sufficient turbulence in the flow field and $Ro \ll 1$ so that the system-scale flow is quasi-two-dimensional \citep{Favier2014, Guervilly2014, Rubio2014, Stellmach2014, Couston2020}. Although our data is suggestive of a domain filling LSV, the limited radial coverage of the chord probe data, $u_c(r > 0.7 R)$, does not allow us to validate the existence of such a structure. Further, this raises the question as to how one would deconvolve a container-scale LSV in a cylindrical domain from a BZF that extends into the fluid bulk.

\section{Discussion \label{sec:discussion}}
We have presented novel simultaneous ultrasound Doppler velocimetry (UDV) and thermal laboratory measurements in rotating liquid metal convection, complemented by direct numerical simulations (DNS). 
The experiments allow an investigation of the flow regime over a wide range of parameters, while the DNS allows a much more detailed analysis of the flow structure at a selected parameter combination.

Our investigation focusses on $Pr \simeq 0.026$ convection in the Ekman number range $4 \times 10^{-5} \leq Ek \leq 5 \times 10^{-6}$ and the Rayleigh number range $9.56\times 10^5 \leq Ra \leq 1.26\times 10^7$. Convection in this system onsets via coherent oscillations inside the bulk of the fluid, and using the onset predictions by \citet{Zhang2009}, our measurements cover supercriticalities $\widetilde{Ra}$ between $1$ and $15.7$. While convective heat transport is relatively inefficient in liquid metals, \JA{the thermal-inertial flow speeds are in good agreement with thermal wind estimates and simultaneously reach about 10$\%$ of free fall velocity $u_{f\!f}$ immediately after the onset of convection and reach up to 50\% for the highest $\widetilde{Ra}$.}  These maximal velocities correspond to nearly 95\% of the RBC flow velocities at the same $Ra$ and $Pr$ number.  Further, the local Rossby number reaches very near unity, $Ro_\ell \simeq 1$, in the $\widetilde{Ra} = 15.7$ case.

Our thermal and UDV data both show that the convective flow becomes quickly multimodal, first via various oscillatory modes, and then at $\widetilde{Ra}>2$, additional wall-attached flow modes are formed in the vicinity of the cylinder side wall. \JA{The wall modes increase the slope of the $Nu$-$Ra$ scaling trend. This increased heat transport efficiency occurs because the wall modes transport temperature anomalies across the entire fluid layer \citep[e.g.,][]{deWit2020, Lu2020}, unlike the oscillatory bulk flows.} The onset frequencies for both the oscillations and the wall modes agree well with the theoretical predictions of \citep{Zhang2009}. The peak oscillatory and wall-mode frequencies and the width of the peaks increases with $\widetilde{Ra}$ in agreement with previous findings by \cite{Horn2017} and \cite{Aurnou2018}. 
\sus{Non-linear wall mode processes lead to the development of a strong sidewall circulation, also known as boundary zonal flow (BZF), well below the stationary onset in our low $Pr$ fluid. The thickness of the BZF appears to converge to $(Ek\, Pr)^{1/3}$, but experiments with varying $Pr$ are required for an unambiguous determination of the scaling properties of the BZF.} At $\widetilde{Ra}>4$, the velocity and temperature spectra indicate broadband turbulence with the vertical velocity field dominated by oscillations even at $\widetilde{Ra}=15.7$. \JA{The bulk vertical and horizontal velocities are comparable in all our experiments, with very near to equipartioned values in the broadband turbulence regime.}

Liquid metal convection is relevant for understanding flows occurring inside Earth's outer liquid metal core \citep{Aurnou2001, Guervilly2016,  Kaplan2017}. Even though convection in spherical shells at low $Pr$ liquid-metal-like fluids onsets via a viscous columnar mode in the equatorial region \citep{Zhang2000, Zhang2017}, recent dynamo models show that at higher supercriticalities the flow inside the tangent cylinder can dominate core dynamics and is characterised by complex, large-scale vortical flows that interact with strongly helical small-scale convection \citep{Schaeffer2017, Cao2018, Aubert2019}. 
The current understanding of the generation of the geomagnetic field is based on convective flows that are dominated by convective Taylor columns that are also found in steady, viscous form in laboratory and numerical experiments in moderate $Pr$ fluids.  Even if equatorial viscous drifting modes characterize the onset of low $Pr$ convection in spherical shells \cite[e.g.,][]{Kaplan2017}, the oscillatory modes may become dominant once convection onsets in the polar regions. Thus, the energetic low $Pr$ oscillatory flows found in this study lead us to question how broadband oscillatory convection in liquid metals will alter the behaviours of global scale dynamos.  

In particular, our DNS results reveal a significant net positive mean helicity in the upper hemicylinder and a negative in the lower hemicylinder.  \JMA{This result supports the conjecture that large-scale dynamo action may be generate by local-scale, low $Pr$ rotating convective flows, as formulated in the multi-scale, asymptotically-reduced model of \citet{Calkins2015}. Our oscillatory rotating convection dominated scenario differs greatly from that of the large eddy dynamo simulations of \citet{Aubert2017}, in which only large-scale ($\ell \gg \ell_O^\infty$) flows exist in the modeled core, and these are singularly responsible for generating planetary magnetic fields over all length and time scales.} 

In addition to the conversion of vertical $rms$ velocity into vertical helicity throughout the fluid bulk, we have also found strong zonal flows focussed in the Stewartson layers, which also extend well into the bulk. Thus, our low $Pr$ rotating convective flows in an upright cylinder contain all the classical kinematic properties that are typically invoked in support of a fully self-consistent dynamo in the oscillatory regime \citep{Moffatt2019}. Future kinematic and fully dynamic dynamo simulations are required to test this hypothesis.

We may also view our cylindrical experiments as a greatly \JMA{oversimplified} model of the fluid within the northern tangent cylinder region of Earth's core \citep{Aurnou2003, Aujogue2018, Cao2018}. Our results suggest that the thermally-driven component of polar convection would tend to break apart into \JA{inertial oscillatory modes, that appear capable of generating significant time-mean helicity \citep[cf.][]{Davidson2015}.} Assuming $Pr \simeq 10^{-2}$ and $Ek \simeq 10^{-15}$ in Earth's core, scaling (\ref{E:Re}) predicts an Earth-like Reynolds number of $Re \sim 10^8$ at a moderately low Rayleigh number of $Ra \simeq 10^{21}$. Furthermore, our results suggest that drifting modes and strong zonal flows might exist along the tangent cylinder \citep[cf.][]{Livermore2017, Aujogue2018, Favier2020}.  The thermo-mechanical boundary conditions, though, differ significantly between our laboratory and DNS cylinders and the tangent cylinder free shear layer in the core.

Having carried out the first thermo-velocimetric study of cylindrical low $Pr$ rotating convection, we envision a number of following studies.  Our experimental range of $1 < \widetilde{Ra} < 16$ was too narrow to build \JA{robust} scaling relations for $Nu$ and $Re$.  Even more so, our local Rossby numbers reached from roughly 0.05 to unity.  Further experiments made over a larger $\widetilde{Ra}$ range will allow for scaling relations to be determined both in the $Ro_\ell \ll 1$ regime and in the $Ro_\ell \gtrsim 1$ regime \JA{\citep{Cheng2018}}. One strength of this study is that it has allowed us to separate oscillatory from stationary modes in the fluid bulk.  \JA{In future studies, we will seek to have both the oscillatory and the stationary bulk modes active simultaneously, in order to determine if one mode is strongly dominant.}

\begin{acknowledgements}
We thank Alex Grannan for support in carrying out these experiments and Gunter Gerbeth, Andr\'e Giesecke, Sven Eckert, \JA{Keith Julien} and Frank Stefani for fruitful discussions and careful reading of the manuscript. Furthermore, we thank gallium for being so much less toxic and less difficult to work with in comparison to its irascible cousins mercury and sodium. TV thanks the Deutsche Forschungsgemeinschaft (DFG) for supporting his work under the grants VO 2331/1, VO 2331/3 and VO 2331/4. JMA and SH thank the NSF Geophysics Program for support via EAR \#1547269 and \#1853196.
\end{acknowledgements}

\bibliographystyle{jfm}
\bibliography{RC_arxiv}

  \begin{center}
	\begin{table}
\def~{\hphantom{0}}
  \begin{tabular}{lcccccccccccc}
    $Ek$ & $Ra$ & $\widetilde{Ra}$ & $Pr$ & $\Omega$ & $P$ & $\Delta T$ & $Nu$ & $Re_{z,max}$ & $u_{f\!f}$ & $u_{z,rms}$ & $u_{z,max}$ & $u_{c,max}$  \\
		 $\times 10^{-5}$ & $\times 10^{6}$ & & & [rad/s] & [W] & [K] & & & [mm/s] & [mm/s] & [mm/s] &  [mm/s] \\[6pt]
4	 & 	0.96	 & 	3.40	 & 	0.027	&	0.42	 & 	50	 & 	3.75	 & 	1.38	 & 	1321	 & 	21.28	 & 	1.5	 & 	4.6	 & 	2.8	\\
4	 & 	1.43	 & 	5.10	 & 	0.027	&	0.42	 & 	100	 & 	5.59	 & 	1.84	 & 	--	 & 	25.97	 & 	--	 & 	--	 & 	--	\\
4	 & 	1.43	 & 	5.10	 & 	0.027	&	0.42	 & 	100	 & 	5.59	 & 	1.84	 & 	1577	 & 	25.97	 & 	1.7	 & 	5.4	 & 	--	\\
4	 & 	2.12	 & 	7.56	 & 	0.027	&	0.42	 & 	200	 & 	8.17	 & 	2.51	 & 	--	 & 	31.39	 & 	--	 & 	--	 & 	--	\\
4	 & 	2.12	 & 	7.56	 & 	0.027	&	0.42	 & 	200	 & 	8.17	 & 	2.51	 & 	2856	 & 	31.39	 & 	2.9	 & 	9.9	 & 	10.6	\\
4	 & 	3.10	 & 	11.04	 & 	0.027	&	0.42	 & 	400	 & 	11.85	 & 	3.47	 & 	3817	 & 	37.82	 & 	4.2	 & 	13.2	 & 	--	\\
4	 & 	3.10	 & 	11.04	 & 	0.027	&	0.42	 & 	400	 & 	11.85	 & 	3.47	 & 	4466	 & 	37.82	 & 	4.8	 & 	15.4	 & 	16.1	\\
4	 & 	4.42	 & 	15.74	 & 	0.027	&	0.42	 & 	800	 & 	16.87	 & 	4.80	 & 	6678	 & 	45.12	 & 	6.8	 & 	23.1	 & 	23.6	\\
2	 & 	1.17	 & 	1.82	 & 	0.027	&	0.84	 & 	50	 & 	4.58	 & 	1.13	 & 	1024	 & 	23.51	 & 	1.0	 & 	3.5	 & 	--	\\
2	 & 	1.97	 & 	3.07	 & 	0.027	&	0.84	 & 	100	 & 	7.63	 & 	1.36	 & 	1676	 & 	30.35	 & 	1.5	 & 	5.8	 & 	4.6	\\
2	 & 	1.99	 & 	3.10	 & 	0.027	&	0.84	 & 	100	 & 	7.70	 & 	1.35	 & 	1588	 & 	30.48	 & 	1.6	 & 	5.5	 & 	--	\\
2	 & 	3.07	 & 	4.80	 & 	0.027	&	0.84	 & 	200	 & 	11.73	 & 	1.75	 & 	2434	 & 	37.62	 & 	2.1	 & 	8.4	 & 	8.6	\\
2	 & 	3.07	 & 	4.79	 & 	0.027	&	0.84	 & 	200	 & 	11.73	 & 	1.75	 & 	2471	 & 	37.62	 & 	2.4	 & 	8.5	 & 	6.5	\\
2	 & 	4.64	 & 	7.25	 & 	0.026	&	0.84	 & 	400	 & 	17.53	 & 	2.35	 & 	4531	 & 	46.00	 & 	4.3	 & 	15.7	 & 	16.0	\\
2	 & 	4.60	 & 	7.18	 & 	0.026	&	0.84	 & 	400	 & 	17.38	 & 	2.37	 & 	4273	 & 	45.79	 & 	3.3	 & 	14.8	 & 	--	\\
2	 & 	6.67	 & 	10.41	 & 	0.026	&	0.84	 & 	800	 & 	25.02	 & 	3.25	 & 	5472	 & 	54.94	 & 	5.7	 & 	18.9	 & 	21.3	\\
2	 & 	6.67	 & 	10.41	 & 	0.026	&	0.84	 & 	800	 & 	25.02	 & 	3.25	 & 	5536	 & 	54.94	 & 	5.8	 & 	19.1	 & 	21.8	\\
1	 & 	1.58	 & 	1.06	 & 	0.027	&	1.71	 & 	60	 & 	6.14	 & 	1.04	 & 	772	 & 	27.22	 & 	0.8	 & 	2.7	 & 		\\
1	 & 	2.33	 & 	1.58	 & 	0.027	&	1.71	 & 	100	 & 	9.01	 & 	1.14	 & 	--	 & 	32.98	 & 	--	 & 	--	 & 		\\
1	 & 	2.34	 & 	1.58	 & 	0.027	&	1.71	 & 	100	 & 	9.05	 & 	1.15	 & 	1276	 & 	33.04	 & 	1.1	 & 	4.4	 & 	1.0	\\
1	 & 	3.46	 & 	2.34	 & 	0.027	&	1.71	 & 	150	 & 	13.16	 & 	1.21	 & 	1639	 & 	39.85	 & 	1.1	 & 	5.7	 & 	--	\\
1	 & 	3.46	 & 	2.34	 & 	0.027	&	1.71	 & 	150	 & 	13.19	 & 	1.19	 & 	2106	 & 	39.89	 & 	1.6	 & 	7.3	 & 	3.8	\\
1	 & 	3.48	 & 	2.35	 & 	0.027	&	1.71	 & 	150	 & 	13.27	 & 	1.19	 & 	1551	 & 	40.01	 & 	1.6	 & 	5.4	 & 	3.1	\\
1	 & 	3.48	 & 	2.35	 & 	0.027	&	1.71	 & 	150	 & 	13.27	 & 	1.19	 & 	1292	 & 	40.01	 & 	1.5	 & 	4.5	 & 	--	\\
1	 & 	4.17	 & 	2.82	 & 	0.026	&	1.71	 & 	200	 & 	15.78	 & 	1.30	 & 	1968	 & 	43.63	 & 	1.7	 & 	6.8	 & 	4.5	\\
1	 & 	6.81	 & 	4.60	 & 	0.026	&	1.71	 & 	400	 & 	25.30	 & 	1.63	 & 	3301	 & 	55.26	 & 	3.1	 & 	11.4	 & 	--	\\
1	 & 	6.87	 & 	4.64	 & 	0.026	&	1.71	 & 	400	 & 	25.51	 & 	1.62	 & 	3247	 & 	55.48	 & 	3.1	 & 	11.2	 & 	13.7	\\
1	 & 	10.2	 & 	6.91	 & 	0.025	&	1.71	 & 	800	 & 	37.41	 & 	2.18	 & 	4987	 & 	67.19	 & 	4.5	 & 	17.2	 & 	--	\\
1	 & 	10.2	 & 	6.93	 & 	0.025	&	1.71	 & 	800	 & 	37.49	 & 	2.17	 & 	5522	 & 	67.25	 & 	4.9	 & 	19.1	 & 	18.4	\\
0.5	 & 	3.01	 & 	0.87	 & 	0.027	&	3.42	 & 	108	 & 	11.53	 & 	1.03	 & 	--	 & 	37.30	 & 	--	 & 	--	 & 		\\
0.5	 & 	3.47	 & 	1.00	 & 	0.027	&	3.42	 & 	127	 & 	13.20	 & 	1.05	 & 	558	 & 	39.92	 & 	0.8	 & 	1.9	 & 		\\
0.5	 & 	4.05	 & 	1.17	 & 	0.027	&	3.42	 & 	150	 & 	15.49	 & 	1.07	 & 	1380	 & 	43.23	 & 	1.4	 & 	4.8	 & 	1.0	\\
0.5	 & 	5.25	 & 	1.52	 & 	0.026	&	3.42	 & 	200	 & 	19.83	 & 	1.10	 & 	1807	 & 	48.92	 & 	1.8	 & 	6.2	 & 	2.0	\\
0.5	 & 	6.63	 & 	1.92	 & 	0.026	&	3.42	 & 	256	 & 	24.86	 & 	1.12	 & 	2108	 & 	54.77	 & 	2.1	 & 	7.3	 & 	3.2	\\
0.5	 & 	7.26	 & 	2.10	 & 	0.026	&	3.42	 & 	300	 & 	27.33	 & 	1.15	 & 	2072	 & 	57.43	 & 	2.1	 & 	7.2	 & 	3.1	\\ \hline 
0.5	 & 	7.71	 & 	2.23	 &  0.026 &	3.42	 & 	325	 & 	28.88	 & 	1.17	 & 	2412	 & 	59.04	 & 	2.2	 & 	8.3	 & 	4.4	\\
\it 0.5  &  \it 8.00     &  \it 2.30 & \it 0.025    &  --       &  --   &   --      & \it 1.21     &  \it 2414     &   \it 60.43  &  \it 2.2  &  \it 8.4 &  \it 5.3\\ \hline
0.5	 & 	8.25	 & 	2.38	 & 	0.026	&	3.42	 & 	340	 & 	30.55	 & 	1.21	 & 	2421	 & 	60.72	 & 	2.1	 & 	8.4	 & 	4.7	\\
0.5	 & 	8.35	 & 	2.41	 & 	0.026	&	3.42	 & 	350	 & 	31.05	 & 	1.21	 & 	1970	 & 	61.21	 & 	2.1	 & 	6.8	 & 	4.1	\\
0.5	 & 	9.10	 & 	2.63	 & 	0.025	&	3.42	 & 	400	 & 	33.22	 & 	1.24	 & 	--	 & 	63.31	 & 	--	 & 	--	 & 	--	\\
0.5	 & 	9.10	 & 	2.63	 & 	0.025	&	3.42	 & 	400	 & 	33.22	 & 	1.24	 & 	2414	 & 	63.31	 & 	2.3	 & 	8.3	 & 	--	\\
0.5	 & 	9.10	 & 	2.63	 & 	0.025	&	3.42	 & 	400	 & 	33.22	 & 	1.24	 & 	--	 & 	63.31	 & 	2.2	 & 	--	 & 	--	\\
0.5	 & 	12.6	 & 	3.65	 & 	0.025	&	3.42	 & 	600	 & 	45.82	 & 	1.40	 & 	3373	 & 	74.36	 & 	3.2	 & 	11.7	 & 	10.4	\\
  \end{tabular}
  \caption{Parameters for the $\Gamma = 2$ laboratory experiments and the $\Gamma = 1.87$ DNS. The first three columns show the derived non-dimensional control parameters: Ekman number $Ek$, Rayleigh number $Ra$ and supercriticality $\widetilde{Ra}=Ra/Ra^{cyl}_{O}$. The next three columns show the measured dimensional control parameters: angular velocity $\Omega$, applied heating power $P$ and vertical temperature difference $\Delta T$. Column 7 and 8 show the Nusselt number $Nu$ and the Reynolds number based on the maximum vertical velocity $Re_{z,max}$. The last four columns show the calculated free-fall velocity $u_{f\!f}$, the UDV root mean square vertical velocity $u_{z,rms}$, the maximum vertical velocities $u_{z,max}$ and chord velocities $u_{c,max}$. Missing values are due to insufficient UDV signal quality. The horizontal lines mark the canonical laboratory case and the DNS case (in italics). The numerical data were rescaled using the material parameters of gallium (see \S 3.1) and the cylinder height $H = 98.4$ mm.}
  \label{tab:kd}
\end{table}
  \end{center}
\end{document}